\pdfoutput=1

\documentclass[11pt]{article}

\usepackage[]{acl}

\usepackage{times}
\usepackage{latexsym}

\usepackage[T1]{fontenc}

\usepackage[utf8]{inputenc}

\usepackage{microtype}

\usepackage{color}
\usepackage{booktabs}
\usepackage{multirow}
\usepackage{graphicx}
\usepackage{array}           
\usepackage{longtable, tabu}  
\usepackage{colortbl} 
\usepackage{tabularx}
\usepackage{enumitem}
\usepackage{amssymb}
\usepackage{subfigure}
\usepackage{amsmath}
\usepackage[nameinlink]{cleveref}
\usepackage{bbding}
\usepackage{pifont}
\usepackage{wasysym}
\usepackage{amssymb}
\usepackage{booktabs}
\usepackage{graphicx}
\usepackage{array}
\crefformat{section}{\S#2#1#3} 
\crefname{algorithm}{Alg.}{Algs.}
\crefformat{subsection}{\S#2#1#3}
\Crefname{equation}{Eq.}{Eqs.}
\Crefname{figure}{Fig.}{Figs.}
\definecolor{Lightgray}{RGB}{110,110,110}

\usepackage{inconsolata}

\title{Investigating Bias in LLM-Based Bias Detection: Disparities between LLMs and Human Perception
}

\author{Luyang Lin$^{1,2}$, Lingzhi Wang$^{3}$\thanks{~~Lingzhi Wang is the corresponding author.}, Jinsong Guo$^{4}$, Kam-Fai Wong$^{1,2}$\\
  $^1$The Chinese University of Hong Kong, China\\
  $^2$MoE Key Laboratory of High Confidence Software Technologies, China\\
  $^3$Harbin Institute of Technology, Shenzhen, China \quad
  $^4$University College London, UK\\
  \tt $^{1,2}$\{lylin, kfwong\}@se.cuhk.edu.hk \\
  \tt $^3$wanglingzhi@hit.edu.cn \quad
    \tt $^4$jinsong.guo@ucl.ac.uk\\
}

\begin{document}
\maketitle
\begin{abstract}
The pervasive spread of misinformation and disinformation in social media underscores the critical importance of detecting media bias. While robust Large Language Models (LLMs) have emerged as foundational tools for bias prediction, concerns about inherent biases within these models persist. In this work, we investigate the presence and nature of bias within LLMs and its consequential impact on media bias detection. Departing from conventional approaches that focus solely on bias detection in media content, we delve into biases within the LLM systems themselves. Through meticulous examination, we probe whether LLMs exhibit biases, particularly in political bias prediction and text continuation tasks. Additionally, we explore bias across diverse topics, aiming to uncover nuanced variations in bias expression within the LLM framework. Importantly, we propose debiasing strategies, including prompt engineering and model fine-tuning. Extensive analysis of bias tendencies across different LLMs sheds light on the broader landscape of bias propagation in language models. This study advances our understanding of LLM bias, offering critical insights into its implications for bias detection tasks and paving the way for more robust and equitable AI systems\footnote{The code is available at \url{https://github.com/lylin0/lin2024investigating}}.
\end{abstract}

\section{Introduction}

Detecting media bias \cite{yu2008classifying,iyyer2014political,liu2022politics} was crucial due to the pervasive spread of misinformation and disinformation on social media platforms, profoundly shaping public perception and decision-making processes. Recently, researchers have increasingly turned to robust LLMs as foundational tools for media bias prediction \cite{lin2024inditag,lin2024indivec,liu2024teller}. Compared to non-pretrained neural models or less powerful language models, LLMs offer enhanced capabilities, yet with an increased risk of bias introduction, given their superior performance and widespread use in media analysis and bias detection. Consequently, there is a growing need to examine bias within the bias detection process itself \cite{fang2023bias,urman2023silence,esiobu2023robbie}.

In this study, we investigate a series of research questions, including whether LLMs exhibit bias, their subsequent impact on media bias prediction results, a fine-grained analysis of LLM bias, and how debiasing affects performance. Before delving into our investigation, it's important to differentiate between the tasks of bias detection and LLM bias analysis. Bias detection in this context pertains to the media bias prediction task, which involves determining whether a given article exhibits bias. This task is text-oriented, focusing on analyzing input text. On the other hand, analyzing bias in LLM involves examining potential biases inherent in the LLM system itself, which is system-oriented and focusing on exploring biases within the system. 

\begin{figure}[t]
\centering
\includegraphics[width=\linewidth]{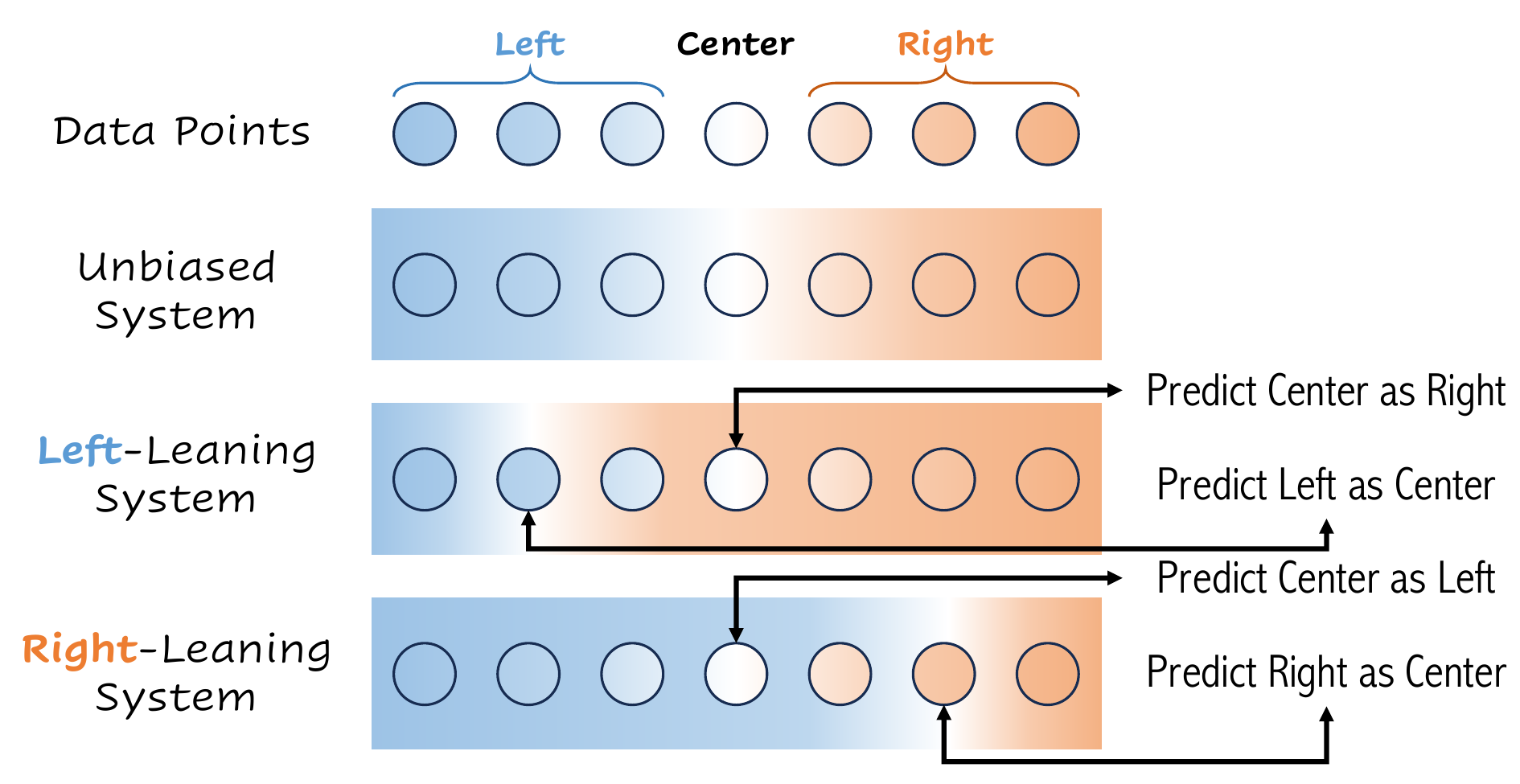}
\vskip -0.5em
\caption{\label{fig:biased_system_interpretation} Interpretation of Biased Systems.}
\vskip -1.5em
\end{figure}
To better illustrate the impact of biased systems on media bias detection, \Cref{fig:biased_system_interpretation} employs political bias prediction as an example. We observe that based on an unbiased system which is capable of accurately predicting the political ideology of given data points, the biased one may exhibit skewed predictions, leading to misinterpretations or misclassifications of the political ideology of the data points. In addition to this illustration, experiments reveal that vanilla GPT-3.5 demonstrates an F1 score of 26.2\% on FlipBias dataset \cite{chen-etal-2018-learning} (a representative political bias prediction dataset), indicating its limited effectiveness in identifying the political leaning of articles. This raises the question of whether the unsatisfactory performance of LLMs in political ideology prediction stems from suboptimal capabilities inherent to LLMs or from inherent biases within the LLMs themselves.

We first explore the research question of whether LLMs exhibit political bias (\colorbox{brown}{\textcolor{white}{RQ1}}) from two distinct perspectives: analyzing LLM bias through political bias prediction and text continuation tasks. The bias prediction perspective enables us to evaluate potential biases in an LLM's comprehension and prediction of specific given articles, while the text continuation perspective offers insights into the political leaning of LLMs' generated content when provided with a short prefix with pre-set political leaning. This yields broader implications of bias in LLMs for  content generation applications.

Furthermore, unlike previous studies \cite{liu2021mitigating,wambsganss-etal-2023-unraveling} that examines bias based on predefined dimensions such as demographics, gender, and location, we aim to explore the bias of LLMs at more granular and flexible levels. This involves examining bias at both predefined and latent topics to address the second research question: \colorbox{brown}{\textcolor{white}{RQ2}} Do LLMs exhibit consistent bias across topics? Further case examination of LLM bias under specific topics and proposed bias evaluation metrics reveal how biases vary across different topics. Through assessing bias consistency across topics that may vary temporally, we gain insights into how LLMs propagate biases.

Furthermore, we explore various debiasing methods, including isolating inherent bias through prompt engineering and adjusting the model's leaning via fine-tuning, to address the question: \colorbox{brown}{\textcolor{white}{RQ3}} How to debias LLMs and further improve performance? Throughout these investigations, we make several key observations that hold significance for future developments in LLM-based frameworks.

Lastly, we assess bias across different LLMs, both open-source and closed-source, to address the fourth research question: \colorbox{brown}{\textcolor{white}{RQ4}} Do various LLMs demonstrate similar bias tendencies? The results suggest that while different LLMs may demonstrate varying bias leanings, bias does indeed exist in the tested LLMs. Moreover, the performance of LLMs does not appear to correlate with the degree of bias exhibited by the models.

In summary, we provide a comprehensive investigation into the presence and nature of bias within LLMs and its consequential impact on media bias detection. The exploration of disparities between LLMs and human perception (i.e., the bias ground truth used in this work is labeled by humans) advances our understanding of LLM bias, offering critical insights into its implications for bias detection tasks and paving the way for more robust and equitable AI systems.

\section{Related Work}

\paragraph{Bias of LMs.}
Understanding bias within LMs is complex due to its normative and subjective nature, often influenced by various contextual and cultural factors \cite{gallegos2023bias}. While providing a formal definition of bias can be challenging, it is commonly observed and studied through its manifestations in LM outputs. Biases manifest in various forms, including representational biases depicting certain social groups negatively \cite{beukeboom2019stereotypes}, disparate system performance leading to misclassifications \cite{blodgett2016demographic}, and reinforcement of normativity \cite{bender2021dangers}. Misrepresentation of social groups can also exacerbate biases \cite{smith2022m}. While research \cite{hada2023fifty,gonccalves2023understanding,conti-wisniewski-2023-using, wang2023causal} has addressed bias in LMs broadly, our work focuses on political standing bias, aiming to elucidate discrepancies between LM cognition and human perceptions.

\paragraph{Bias Mitigation.} Bias mitigation techniques encompass pre-processing, in-training, intra-processing, and post-processing interventions \cite{gallegos2023bias}. Pre-processing involves altering model inputs, such as data and prompts \cite{venkit2023nationality}, to create more representative training datasets through techniques like data augmentation \cite{qian2022perturbation}, data filtering \cite{garimella2022demographic}, prompt modification \cite{venkit2023nationality}, and debiasing pre-trained representations. Intra-processing methods \cite{zayed2023deep} modify model behavior at inference without further training, including decoding strategies, post hoc model adjustments, and modular debiasing networks. In-training techniques aim to reduce bias by modifying the optimization process, such as adjusting loss functions \cite{liu2021mitigating}, updating probabilities, freezing parameters \cite{gira2022debiasing}, or neuron removal \cite{joniak2022gender} during training. Post-processing \cite{tokpo2022text} mitigates bias in model outputs through techniques like identifying and replacing biased tokens without altering original model parameters.

\section{RQ1: Do LLMs exhibit political bias?}\label{sec:rq1:llm_left_leaning}
Previous work \citet{rozado2023political} conducted 15 different political orientation tests on ChatGPT. The findings reported by \citet{rozado2023political} reveal that ChatGPT tends to exhibit a preference for left-leaning viewpoints in its responses to questions. However, it is noteworthy that their investigations were based on a limited number of political orientation tests (i.e., 15 tests). In this section, we employ various bias analysis methods to further investigate the political bias exhibited by LLMs.

\subsection{LLM-based Bias Prediction}\label{ssec:rq1_prediction}
\begin{table*}[t]
\setlength{\tabcolsep}{1mm}\small
\begin{center}
\begin{minipage}[t]{0.68\linewidth}
\hspace{-0.04\linewidth} 
\resizebox{\linewidth}{!}{
\begin{tabular}{p{2cm}|ccc|c|c|p{2.5cm}}
\toprule[1.0pt]
Dataset & \multicolumn{3}{c}{Bias Label}  & \# of Instances & Avg Length & Source \\
\midrule[0.5pt]
 \multirow{2}{\linewidth}{FlipBias \cite{chen-etal-2018-learning}} & Left & Center & Right & 3,066 & 1,077 & \multirow{3}{\linewidth}{New York Times, Huffington Post, Fox News and Townhall} \\
 & 33.3\% & 33.3\% & 33.3\% & & & \\
\cmidrule(lr){1-6}
 \multirow{2}{\linewidth}{ABP \cite{baly2020we}} & Left & Center & Right & 37,554 &  1,095 &  \\
 & 34.5\% & 36.6\% & 28.8\% & & & \\
\bottomrule[1.0pt]
\end{tabular}
}
\captionof{table}{\label{tab:dataset_statistic} Statistics of FlipBias and ABP.}
\end{minipage}%
\hspace{0.02\linewidth} 
\begin{minipage}[t]{0.22\linewidth}
\vspace{-4\abovecaptionskip}
\includegraphics[width=\linewidth]{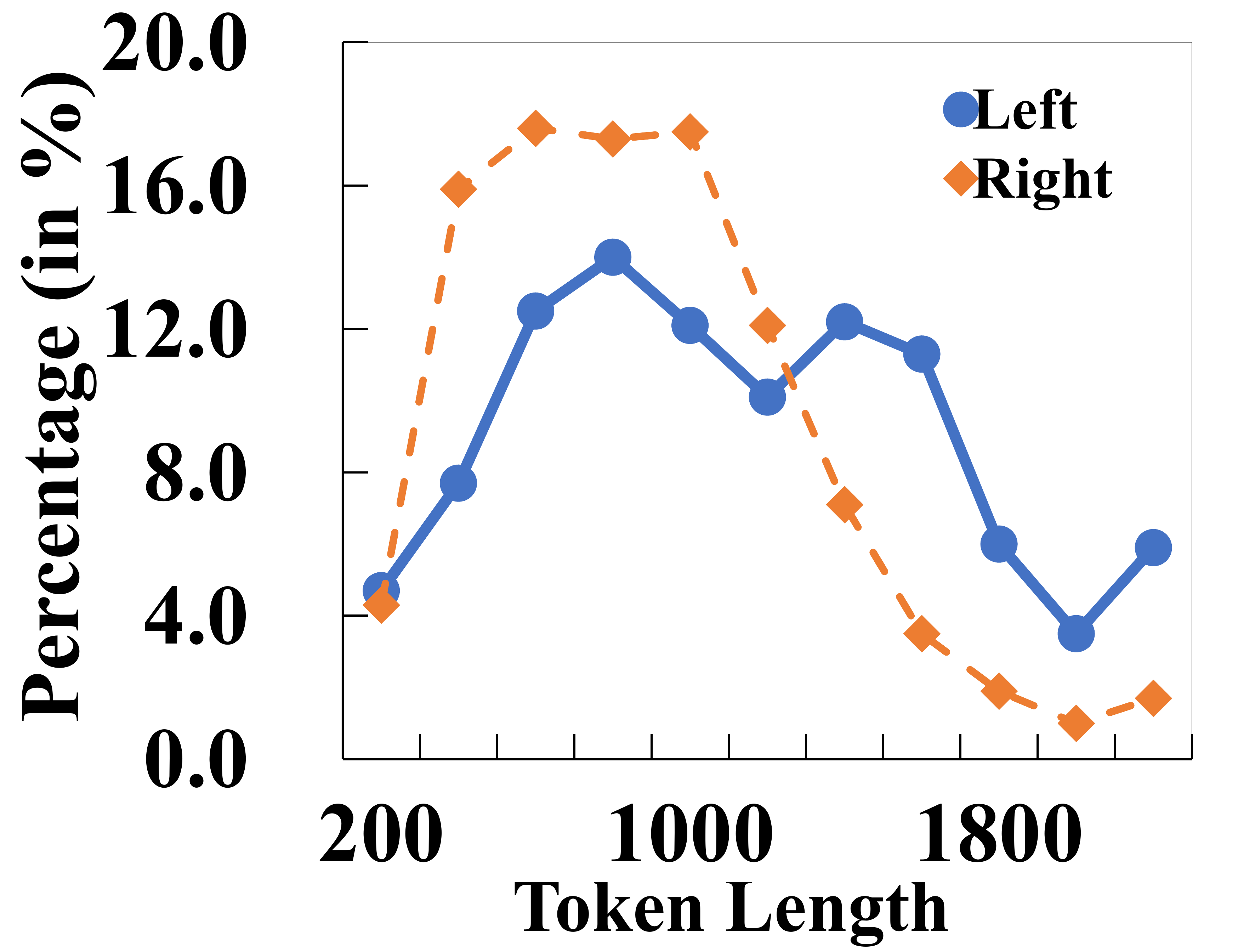}
\captionof{figure}{\label{fig:flipbias_len_distribution}FlipBias Len.}
\end{minipage}
\end{center}
\vskip -1.5em
\end{table*}

\begin{figure}[t]
\centering
\subfigure[Political Bias Prediction on FlipBias] {\label{sfig:fb_ground_predicted}
\includegraphics[width=0.8\linewidth]{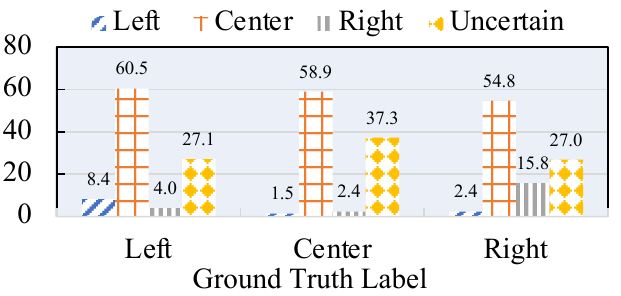}
}
\subfigure[Political Bias Prediction on ABP] {\label{sfig:abp_ground_predicted}
\includegraphics[width=0.8\linewidth]{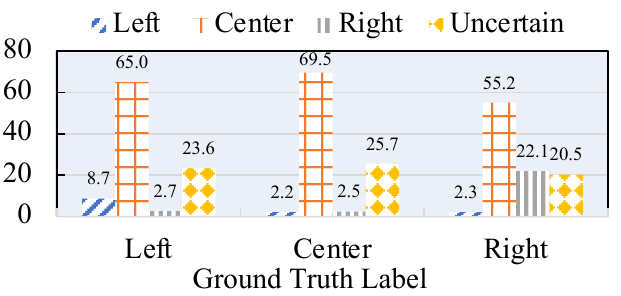}
}
\vskip -1em
\caption{\label{fig:FB_ABP_ground_predicted} LLM's prediction on FlipBias and ABP.
}
\vskip -1em
\end{figure}
We adopt vanilla ChatGPT model to conduct political leaning prediction on two popular datasets (i.e., FlipBias \cite{chen-etal-2018-learning} and ABP \cite{baly2020we}). The statistic of these two datasets can be found in \Cref{tab:dataset_statistic}. We can see that there are 1022 triples (i.e., each triple is with left-, center-, right-leaning article on the same event) in FlipBias and more than 30k instances in ABP dataset. For each instance, we prompt  \texttt{gpt-3.5-turbo-0613} with the following instruction to get the bias prediction results of vanilla ChatGPT:
\vskip -0.5em
\begin{quote}
\setlength\leftskip{-0.4cm}
\textit{Given the article provided below: \colorbox{Lightgray}{\color{white}{\textit{TEXT ARTICLE}}} \\
Analyze the text content and assign a label from \{left, right, center, uncertain\}. In this context, ``left'' indicates a left-leaning article, ``right'' signifies a right-leaning article, ``center'' implies no obvious political leaning, and ``uncertain'' denotes that the political orientation could not be determined. Please provide your analysis and output a new single line containing only the assigned label.
}
\end{quote}
\vskip -0.5em
We present the bias prediction results in \Cref{fig:FB_ABP_ground_predicted}, comparing the ground truth labels (left, center, right) with the model's predictions (left, center, right, uncertain). Before delving into the analysis of the results in \Cref{fig:FB_ABP_ground_predicted}, we establish the following assumption. $\mathcal{A}0$: LLMs exhibit inherent political cognitive bias rather than an overall inability to judge articles' political leaning. $\mathcal{A}0$ implies that the prediction results of LLMs follow a linear bias pattern, as illustrated in \Cref{fig:biased_system_interpretation}. Based on the results in \Cref{fig:FB_ABP_ground_predicted}, we have the following observations: \\
$\bullet$~ \textit{$\mathcal{O}1$: The tested LLM exhibits left-leaning viewpoints.} By focusing on the proportions of Left-Center (where Left is the ground truth label and Center is the predicted label, e.g., the Left-Center proportion in \Cref{sfig:fb_ground_predicted} is 60.5) and Right-Center presented in \Cref{fig:FB_ABP_ground_predicted}, we observe that the Left-Center values surpass the Right-Center values on both datasets. These higher values indicate that the tested LLM demonstrates a left-leaning political cognitive bias, resulting in a higher likelihood of predicting left-leaning articles as centered articles. Furthermore, by comparing the Center-Left and Center-Right values across two datasets, we observe that the tested LLM tends to predict the centered article slightly more as right-leaning rather than left-leaning. This observation is consistent with the notion that the tested LLM exhibits left-leaning viewpoints.\\
$\bullet$~ \textit{$\mathcal{O}2$: Despite left-leaning tendencies, the tested LLM excels in predicting right-grounded articles.} An examination of the proportions of Left-Left and Right-Right predictions in \Cref{fig:FB_ABP_ground_predicted} reveals that the Right-Right proportions are significantly higher than those of Left-Left. This suggests that the tested LLM excels in accurately classifying articles with a right-leaning perspective. \\

By comparing the results predicted by LLMs, we derive initial observation $\mathcal{O}1$, which is consistent with the findings reported by \citet{rozado2023political}. In the following, we explore the viewpoint leaning of LLMs through Article Continuation experiments and two distinct analytical approaches.

\subsection{LLM-based Article Continuation}
\label{ssec:article_continuation}
Beyond the prediction-based analysis outlined earlier, we investigate LLM bias through article continuation. By supplying LLMs with prefixes derived from political articles and prompting them to extend these prefixes, we assess the political leaning of the generated suffix to analyze the inherent bias of LLMs. Our evaluation employs two methods for determining the political leaning of generated content: intuitive embedding-based similarity matching and Left and Right Vocabulary-based matching, following the approach proposed by \cite{fang2023bias, wambsganss2023unraveling}.

Following this, we begin by providing a detailed description of the continuation implementation and then proceed to conduct in-depth examinations of bias in LLMs based on two distinct methods for determining political leaning of continued content.

\paragraph{Article Continuation.} We prompt \texttt{gpt-3.5-turbo-0613} with a continuation prompt to generate text based on the given prefix.
\begin{quote}
\textit{Continue the text provided below: \colorbox{Lightgray}{\color{white}{\textit{TEXT ARTICLE}}}}
\end{quote}
Building on the core idea of assessing the generated suffix to reflect the leaning of LLMs, we explore two automated methods to determine the bias label of the generated content.

\paragraph{Embedding-Based Similarity Matching.} We utilize an off-the-shelf text embedding API of ChatGPT to create a vector database following \cite{peng2023embedding}. Specifically, the vector database comprises embeddings of all instances from the FlipBias dataset. For each instance in the FlipBias dataset, we construct prefixes (e.g., prefixes with a fixed number of tokens such as 20, 40, etc.) and obtain the continued suffix by prompting ChatGPT with the previously introduced prompt. Subsequently, we label the continuation suffix by calculating the similarity between the generated suffix and tripled instances\footnote{The triples are adjusted to match the length of the prefix, considering a prefix of length $n$, resulting in a length minus $n$.} (i.e., left-leaning, center-leaning, and right-leaning articles) centered around the same event. We determine the bias label of the generated text based on the label of the instance with the highest similarity score. 
The entire process is formally described as follows.
\begin{align}
    \text{Similarity}_i &= \frac{v_{\text{suffix}} \cdot v_i}{|v_{\text{suffix}}||v_i|}, \quad i \in \{\text{left, center, right}\} \\
    \text{Bias Label} &= \text{argmax}(\text{Similarity}_i)
\end{align}
\noindent where $v_{(.)}$ represents the embedding of the text.

\paragraph{Left and Right Vocabulary-Based Matching.} By following \citet{yano2010shedding}, we first construct two vocabularies for left-and right-leaning articles separately. Each vocabulary is constructed by doing statistic of the word frequency for articles with ground-truth left and right labels and removing stop words (details are shown in \Cref{appendix:sec:left_right_vocabulary_construction}), which can represent the characteristic of the respective political leaning. The presence of a higher number of words from a specific vocabulary within an article indicates the alignment of the article with the corresponding political leaning. For instance, an article featuring more tokens from the left-leaning vocabulary indicates its left-leaning orientation.
\begin{figure}[t]
\centering
\includegraphics[width=0.85\linewidth]{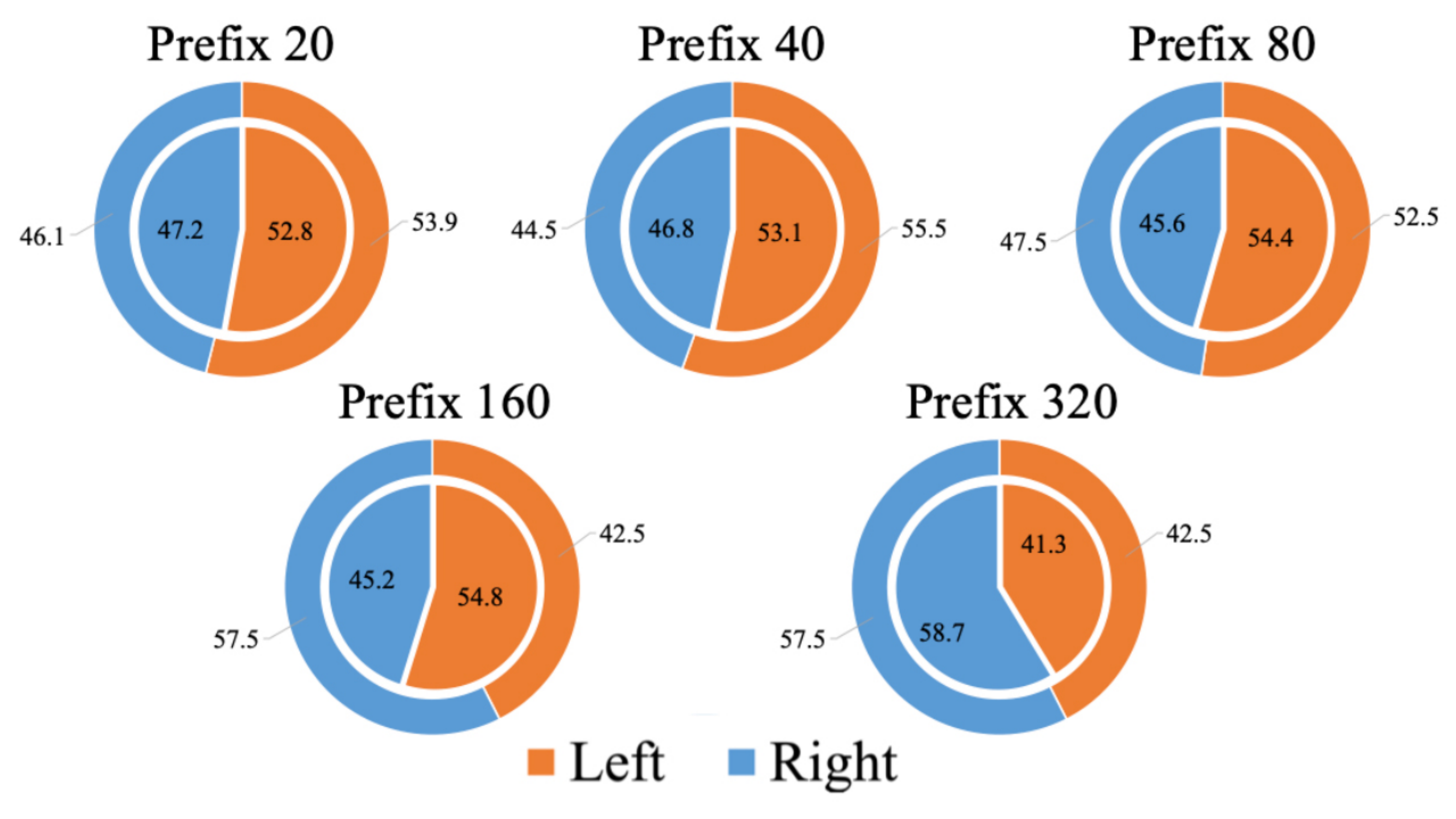}
\vskip -0.5em
\caption{\label{fig:article_contiuation_results} Article Continuation Results on FlipBias: The inner pie chart presents the outcomes of embedding-based similarity matching, while the outer doughnut illustrates the results of vocabulary-based matching.}
\vskip -0.5em
\end{figure}

In \Cref{fig:article_contiuation_results}, we present the outcomes of article continuation experiments with varying prefix lengths (e.g., 20, 40 tokens) employing both embedding-based and vocabulary-based matching.  It's important to note that only the relative percentages of left and right are presented, disregarding the center situation. From \Cref{fig:article_contiuation_results}, we can see that across prefix lengths ranging from 20 to 80, both label matching methods consistently show a higher percentage of left predictions, suggesting a left-leaning trend in continued articles. However, as the prefix length increases to 320, both methods begin to predict continued articles as more right-leaning. This change may be attributed to the fact that the average length of right-leaning articles is shorter than left-leaning articles (refer to \Cref{fig:flipbias_len_distribution}). 
Additionally, an analysis of left-leaning articles reveals that sentences with left-leaning bias typically appear in the latter part of the article and rarely appear in the prefix.
Therefore, when given a prefix with 320 tokens, the political leaning of the prefix becomes clearer, representing a substantial portion—approximately 40\%—of the average length of Right articles (794 tokens) and 28\% of Left articles (1111 tokens). This clearer representation of political leaning in the prefix makes it more likely for the LLM to generate a right-leaning suffix. Consequently, LLMs may find it easier to predict right-leaning continued suffixes.

\paragraph{Discussion about Results of Classifier}
We also conduct the experiments based on the classifier \texttt{gpt-3.5-turbo-0613}. We observe that when we use the vanilla ChatGPT model, the model predicts about $70\%$ of the generated articles as Center, compared to around $15\%$ of the articles matched as Center in our other methods. The model prefers to regard the text generated by itself as objective. It needs to be further considered whether we can use a classifier that may contain bias to measure the bias. When we give the model three articles of the same event as references (the three articles are the same as we use in embedding-based similarity matching), the trends are similar to the results of embedding-based similarity. More details and results are in \Cref{appendix:ssec:classifier}.
\section{RQ2: Do LLMs demonstrate consistent bias across all topics?}
\begin{table}[t]
\setlength{\tabcolsep}{1mm}\small
\begin{center}
\begin{tabular}{c|ccc}
\toprule[1.0pt]
Dataset &\# of Topics & Latent & Avg Instance \# Per Topic  \\
\midrule[0.5pt]
FlipBias & 152 &\CheckmarkBold & 82 \\
\midrule
ABP & 108 &\XSolidBrush & 348 \\
\bottomrule[1.0pt]
\end{tabular}
\end{center}
\vskip -1em
\caption{\label{tab:topic_statistic} Statistics of Topics in FlipBias and ABP.
}
\vskip -1em
\end{table}
\begin{figure}[t]
\centering
\subfigure[Topic LCRC on FlipBias] {\label{sfig:FB_LCRC}
\includegraphics[width=0.45\linewidth]{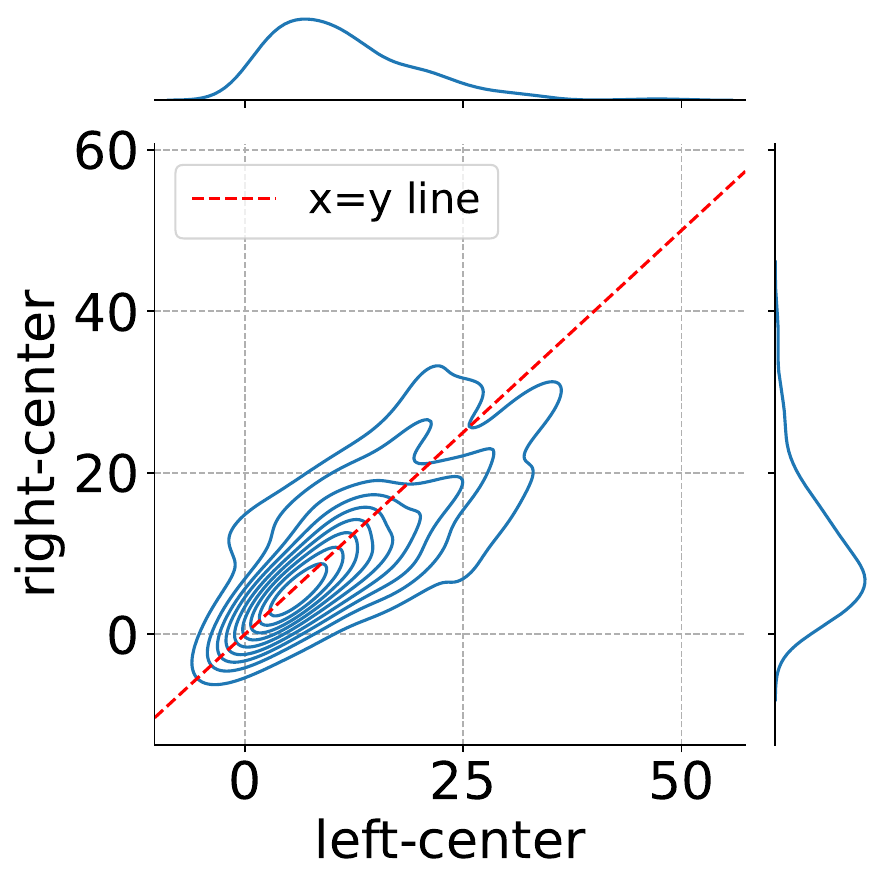}
}
\subfigure[Topic CLCR on FlipBias] {\label{sfig:FB_CLCR}
\includegraphics[width=0.45\linewidth]{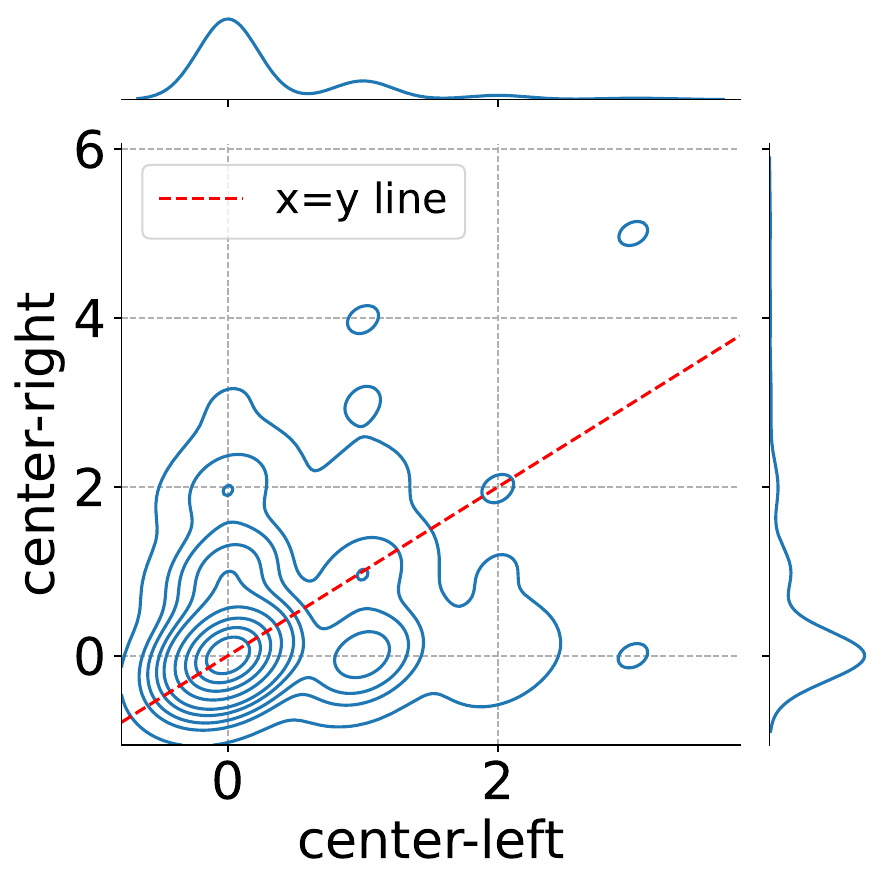}
}
\subfigure[Topic LCRC on ABP] {\label{sfig:ABP_LCRC}
\includegraphics[width=0.45\linewidth]{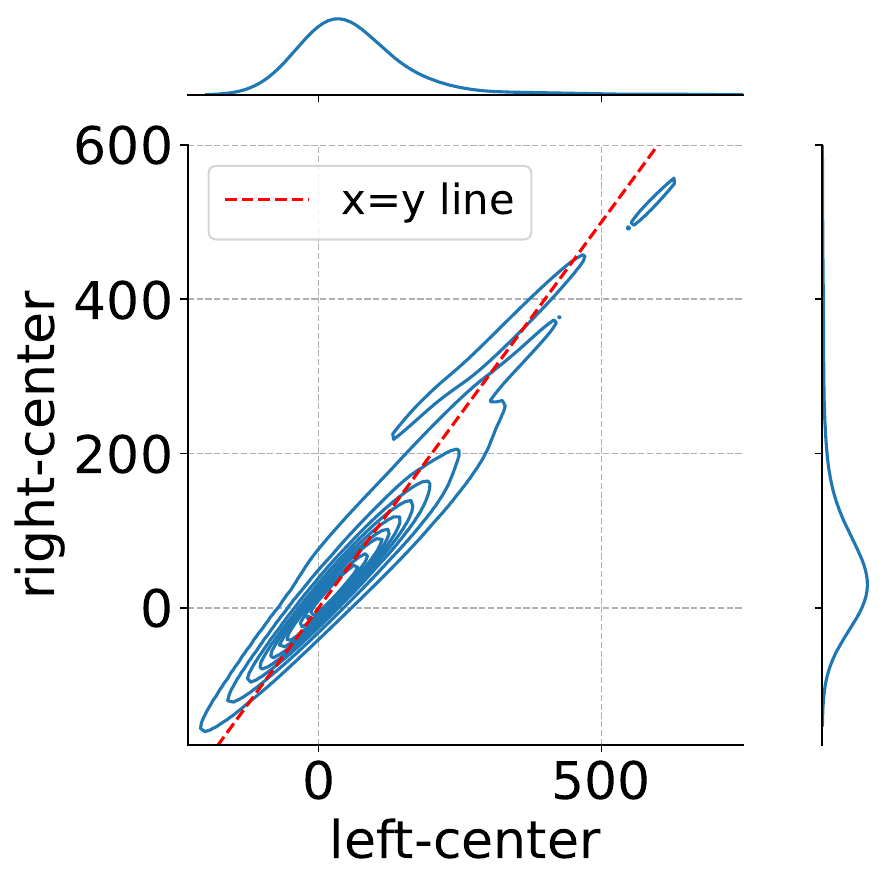}
}
\subfigure[Topic CLCR on ABP] {\label{sfig:ABP_CLCR}
\includegraphics[width=0.45\linewidth]{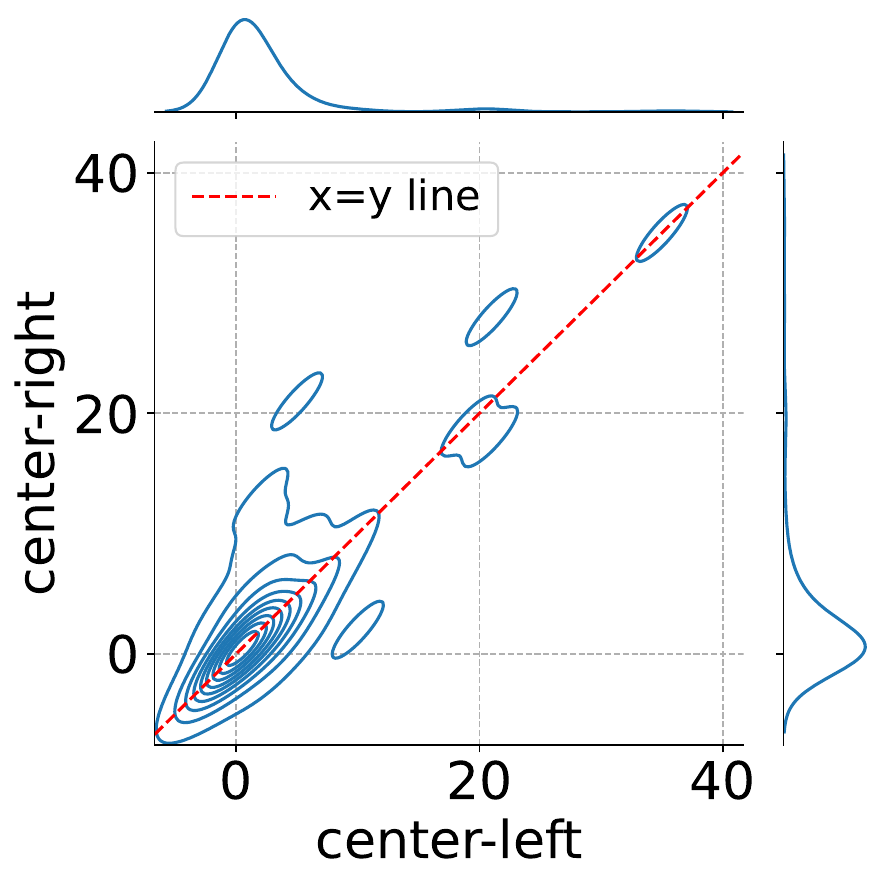}
}
\vskip -1em
\caption{\label{fig:FB_ABP_LCRC_CLCR} Joint plot displaying kernel density estimates.
}
\vskip -1em
\end{figure}

As elaborated in \Cref{ssec:rq1_prediction}, our tested LLM exhibits a left-leaning bias compared to viewpoints derived from the ground-truth labels assigned by human evaluators. In this section, we delve into whether the LLM consistently showcases a leaning across all discussed topics. While the ABP dataset includes topic information, the FlipBias dataset lacks such information inherently. To address this, we construct latent topics following the methodology proposed by \cite{lin2024indivec}. The detailed process of latent topic construction is provided in \Cref{appendix:sec:latent_topic_construction}. As the constructed latent topics of FlipBias dataset are not predefined, we attempt to demonstrate their relevance and coherence to predefined topics in ABP dataset. This is achieved by presenting statistics on the (latent) topics of both datasets in \Cref{tab:topic_statistic}, and by plotting joint distributions of Left-Center (i.e., where the ground-truth label is left and the predicted label is center) and Right-Center, as well as Center-Left and Center-Right, accompanied by kernel density estimates in \Cref{fig:FB_ABP_LCRC_CLCR}.  
To interpret the figure, points farther from the origin indicate more incorrect judgments, while deviation from the $x=y$ line reflects an imbalance between cases classified as Right and Left.
It is evident that the joint plots of the FlipBias and ABP datasets exhibit similar patterns. The main difference arises in the distributions based on predefined topics (i.e., \Cref{sfig:ABP_LCRC} and \Cref{sfig:ABP_CLCR}), which appear more focused compared to the distributions based on latent topics (i.e., \Cref{sfig:FB_LCRC} and \Cref{sfig:FB_CLCR}), which demonstrate greater dispersion.

\paragraph{Visualization Based on Bias Tendency Index.} Before presenting the results of viewpoints leaning in LLMs, we introduce two Bias Tendency Index (BTI) as follows.

\begin{equation}\small\label{eq:bti-1}
    \text{BTI-1} = \frac{\text{Count(left-center)}}{\text{Count(left)}} -
\frac{\text{Count(right-center)}}{\text{Count(right)}} 
\end{equation}
\begin{equation}\small\label{eq:bti-2}
    \text{BTI-2} = \frac{\text{Count(center-right)}}{\text{Count(center)}} -
\frac{\text{Count(center-left)}}{\text{Count(center)}} 
\end{equation}

\noindent where BTI-1 measures the bias tendency of the tested LLM regarding left and right-ground truth labeled articles. It quantifies the difference in predicting articles as center when the ground truth is left versus right. Similarly, BTI-2 focuses on the bias tendency of the LLM concerning articles with a ground truth label of center. It measures the disparity in predicting articles as right or left when the ground truth is center. A positive BTI-1 (BTI-2) suggests the tested LLM shows a left-leaning viewpoints, while a negative value suggests a right-leaning bias of LLM. 

\begin{figure}[t]
\centering
\subfigure[Distribution on FlipBias] {\label{sfig:fb_topic_vi-v2}
\includegraphics[width=0.45\linewidth]{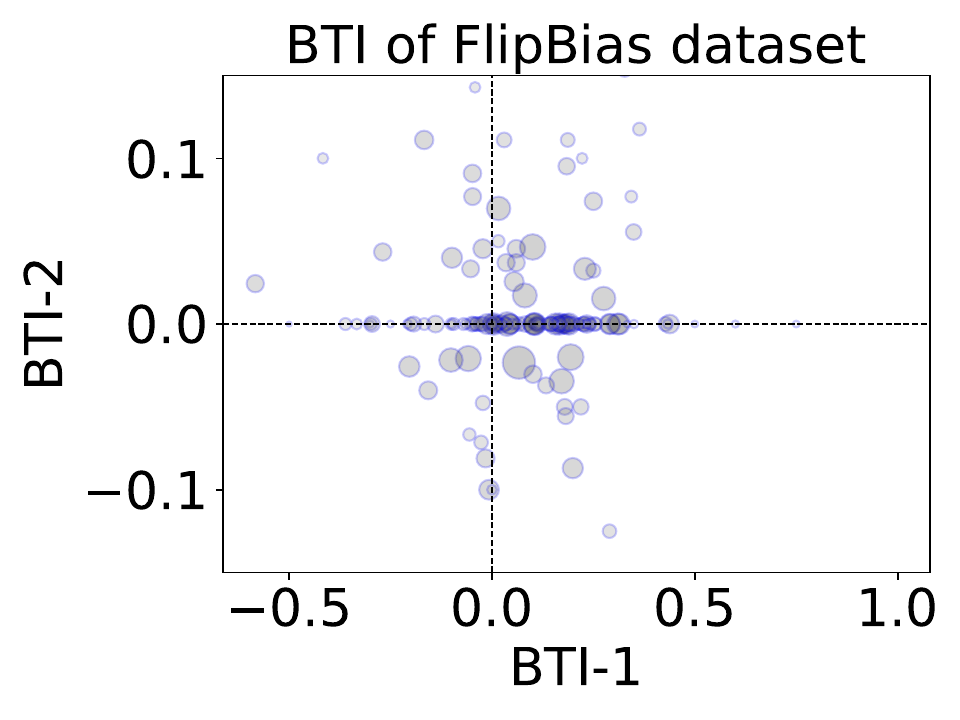}
}
\subfigure[Distribution on ABP] {\label{sfig:abp_topic_vi-v2}
\includegraphics[width=0.45\linewidth]{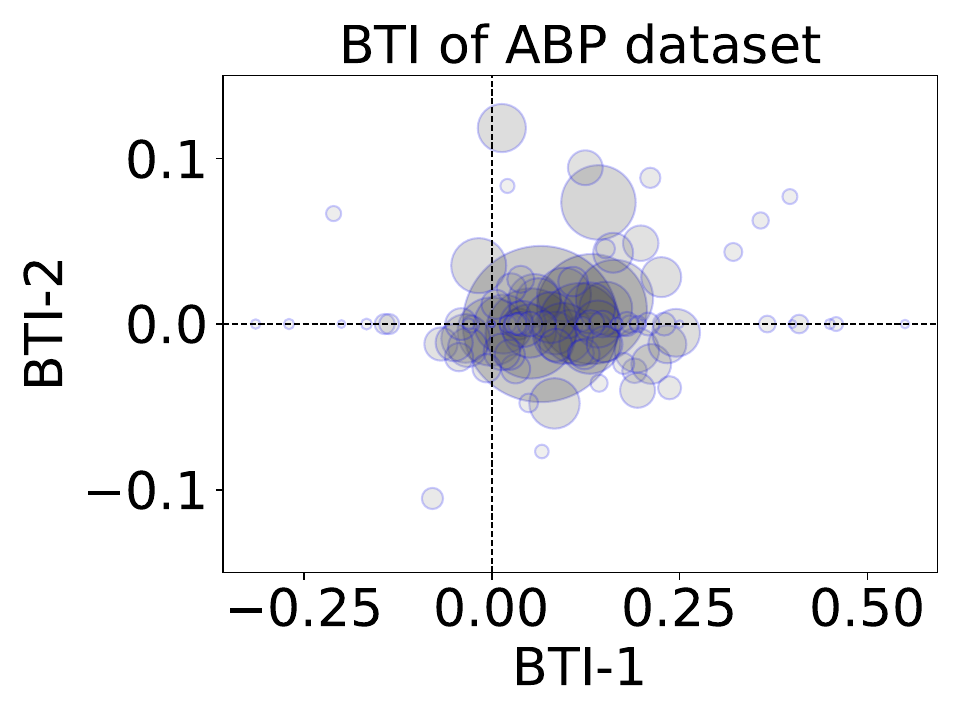}
}
\vskip -1em
\caption{\label{fig:FB_ABP_topic_vi-v2} BTI distribution of Topics on FlipBias and ABP. Darker colors and larger circles indicate more instances under the corresponding topic.
}
\vskip -1em
\end{figure}
We present the distribution of BTI-1 and BTI-2 for the FlipBias and ABP datasets in \Cref{fig:FB_ABP_topic_vi-v2}. Each point in \Cref{fig:FB_ABP_topic_vi-v2} represents a distinct topic, larger points indicate more instances located in the corresponding topic, and darker regions imply more topics located in that region. From \Cref{fig:FB_ABP_topic_vi-v2}, we find:\\
$\bullet$~ \textit{$\mathcal{O}3$: The tested LLM does not exhibits same viewpoint leaning on all topics.} As discussed in \Cref{sec:rq1:llm_left_leaning} (i.e., $\mathcal{O}1$), the tested LLM demonstrates an overall left-leaning viewpoint on both the Flipbias and ABP datasets. By presenting the BTI-1 and BTI-2 values (where a positive value indicates left-leaning, referring to the explanation to \Cref{eq:bti-1} and \Cref{eq:bti-2}) for all topics in Figure 4, it is evident that while most points are situated in the right region of the figure (i.e., BTI-1 $> 0$), there are topics with notably negative values, indicating that the tested LLM displays right-leaning viewpoints on these topics.

$\bullet$~ \textit{The distribution of BTI-1 is more pronounced compared to the BTI-2 value}. Both \Cref{sfig:fb_topic_vi-v2} and \Cref{sfig:abp_topic_vi-v2} exhibit clear left-leaning tendencies in the distribution of BTI-1. While the distribution of BTI-2 on these two datasets appears more evenly spread, with points displaying both negative and positive BTI-2 values generally at similar scales.

$\bullet$~ \textit{The topic frequency distribution on FlipBias appears more evenly distributed compared to that of the ABP dataset.} By examining the sizes of points in \Cref{sfig:fb_topic_vi-v2} and \Cref{sfig:abp_topic_vi-v2}, it is apparent that the clustered latent topics of FlipBias are more evenly distributed, indicating a balanced number of instances contained within each cluster. We provide interpretations of some clustered latent topics and the contained indicators in \Cref{appendix:sec:latent_topic_construction}.

\paragraph{Case Study of Biased Topics.} 
\begin{figure}[t]
\centering
\includegraphics[width=\linewidth]{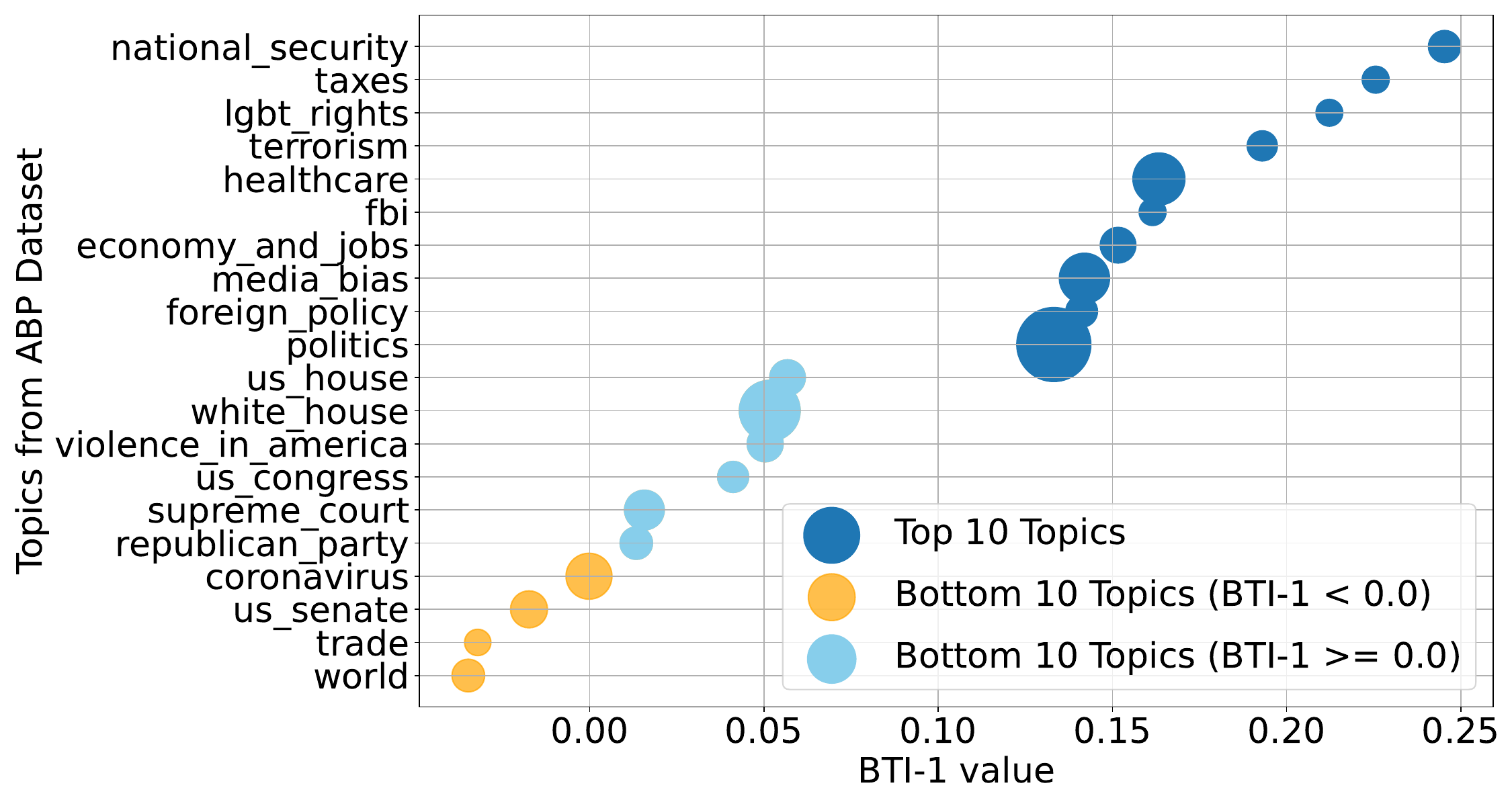}
\vskip -0.5em
\caption{\label{fig:abp_topic_case_study_plot} BTI-1 distribution for Top and Bottom 10 topics (ranked by BTI-1) with above-average frequency (i.e., the number of instances under respective topic).}
\vskip -0.5em
\end{figure}
\newcommand{\colorcell}[2]{\cellcolor{#1!#210}}

\begin{table*}[t]
\setlength{\tabcolsep}{1mm}\small
\begin{center}
\begin{tabular}{p{12cm}|ccc}
\toprule[1.0pt]
Interpretation of Top and Bottom 5 latent
topics (ranked by BTI-1 values)  &BTI-1& BTI-2& Frequency  \\
\midrule[0.5pt]
Comprehensive Use of Quotes and Citations in Journalism &\cellcolor{blue!44}0.44&0.00& 81 \\
Diverse Perspectives on President Obama's Policies and Actions &\cellcolor{blue!31}0.31&0.00& 99  \\
Analysis of Recent Terrorist Attacks and Security Measures in Various Cities &\cellcolor{blue!31}0.31&0.00& 103 \\
Critique of DACA Amnesty Program and Advocacy for Stricter Immigration Policies &\cellcolor{blue!29}0.29&0.00& 89 \\
Diverse Rhetorical Strategies in Political Discourse &\cellcolor{blue!29}0.29&0.00& 97 \\
\midrule
Trump's Clashes with Federal Law Enforcement and Media &\cellcolor{orange!17}-0.17&0.11& 80 \\
Analysis of Media Coverage Surrounding Trump's Ratings, Criticisms, and Mental Fitness &\cellcolor{orange!20}-0.20&-0.03& 98 \\
Trump Administration's Response to Russia Sanctions and Political Fallout &\cellcolor{orange!27}-0.27&0.04& 71 \\
Satirical Commentary and Critique on Political Events and Figures &\cellcolor{orange!30}-0.30&0.00& 58 \\
Media Coverage of Trump Administration &\cellcolor{orange!58}-0.58&0.02& 70 \\
\bottomrule[1.0pt]
\end{tabular}
\end{center}
\vskip -1em
\caption{\label{tab:fb_topic_case_study} Interpretation of Top and Bottom 5 Latent Topics on FlipBias. When interpreting BTI-1 and BTI-2 values, a BTI value closer to zero indicates less bias in the model. Positive values suggest a left-leaning bias, while negative values indicate a right-leaning bias.}
\vskip -1em
\end{table*}

To further analyze the LLM's leaning across various topics, we utilize several cases from the FlipBias and ABP datasets to demonstrate the relationship between viewpoint leaning and topic. For a more representative analysis, we select topics with above-average frequency and then rank them based on the calculated BTI-1 values. We present the top 5 and bottom 5 latent topics from FlipBias in \Cref{tab:fb_topic_case_study}. The interpretation of latent topics is obtained by prompting ChatGPT to provide a summary of the cluster indicators. More latent topic cases ranked by BTI-2 values can be found in \Cref{appendix:sec:latent_topic_construction}.

From \Cref{tab:fb_topic_case_study}, we observe that the trend of BTI-2 values is more centered around $0.0$ when the range of BTI-2 extends to $\pm0.5$, which is consistent with the observation of \Cref{fig:FB_ABP_topic_vi-v2}. 
The LLM's left-leaning viewpoints on topics (upper part of \Cref{tab:fb_topic_case_study}) like journalism's use of citations, Obama's policies, and immigration critique reflect values of transparency, inclusivity, and social justice. This aligns with the narrative often seen in left-leaning media, emphasizing fact-checking, diverse perspectives, and human rights advocacy. These viewpoints may shaped by the model's training data and structural biases.
The prevalence of Trump-related topics among the bottom 5 latent topics (lower part of \Cref{tab:fb_topic_case_study}) with negative BTI-1 suggests a potential right-leaning bias in the language model's treatment of Trump administration subjects. Given FlipBias's data collection primarily from 2013 to 2018, a period marked by heightened political polarization, this alignment hints at a correlation between temporal context and exhibited biases.

We further plot the BTI-1 distribution of the Top and Bottom 10 topics (ranked by BTI-1 values) with above-average frequency for the ABP datasets in \Cref{fig:abp_topic_case_study_plot}. 
Upon closer examination, notable similarities emerge between topics with extreme values in both the Flipbias and ABP datasets. The analysis reveals similarities between extreme value topics in both Flipbias and ABP datasets, with positive values often focusing on security and terrorism, and negative values frequently discussing Trump's government and the US-China trade war. Given that ABP dataset's data is collected between 2019-2020, we infer that short-term hot topics like coronavirus tend to exhibit negative bias, while broader subjects like LGBT rights trend positively. The concentration of articles in the middle range of topics suggests that data scale may influence bias trends, with widely discussed topics reflecting human perspectives more closely. 

\section{RQ3: How to debias LLMs and further improve performance?}
\begin{figure*}[t]
\centering
\subfigure[BLE] {\label{sfig:FB_debias_topic_v1_v2_BLE}
\includegraphics[width=0.225\linewidth]{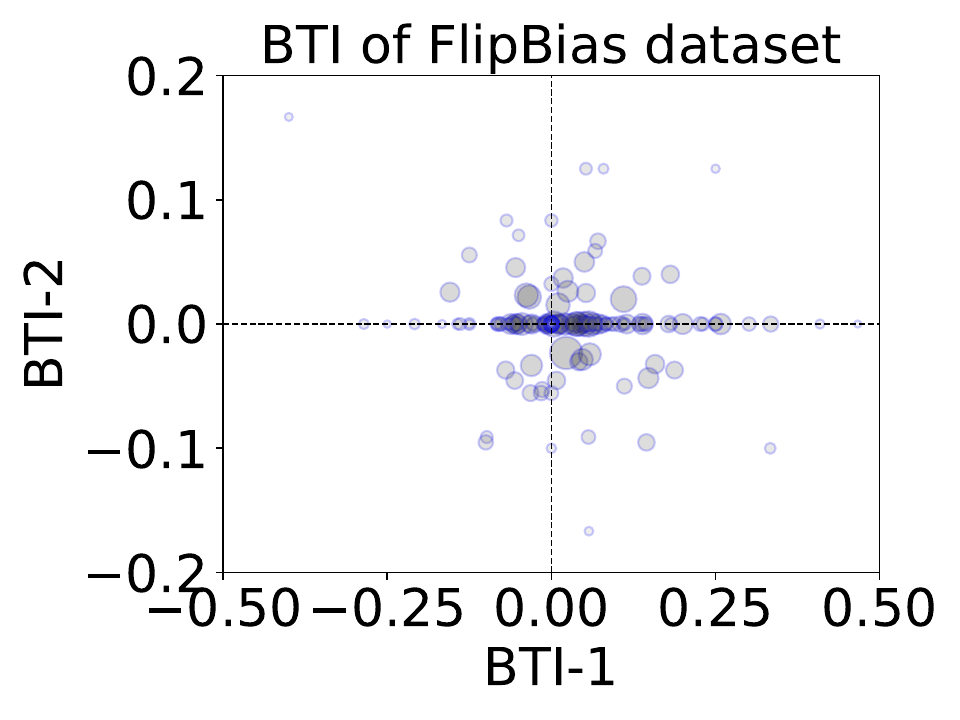}
}
\subfigure[3-Shot] {\label{sfig:FB_debias_topic_v1_v2_3shot}
\includegraphics[width=0.225\linewidth]{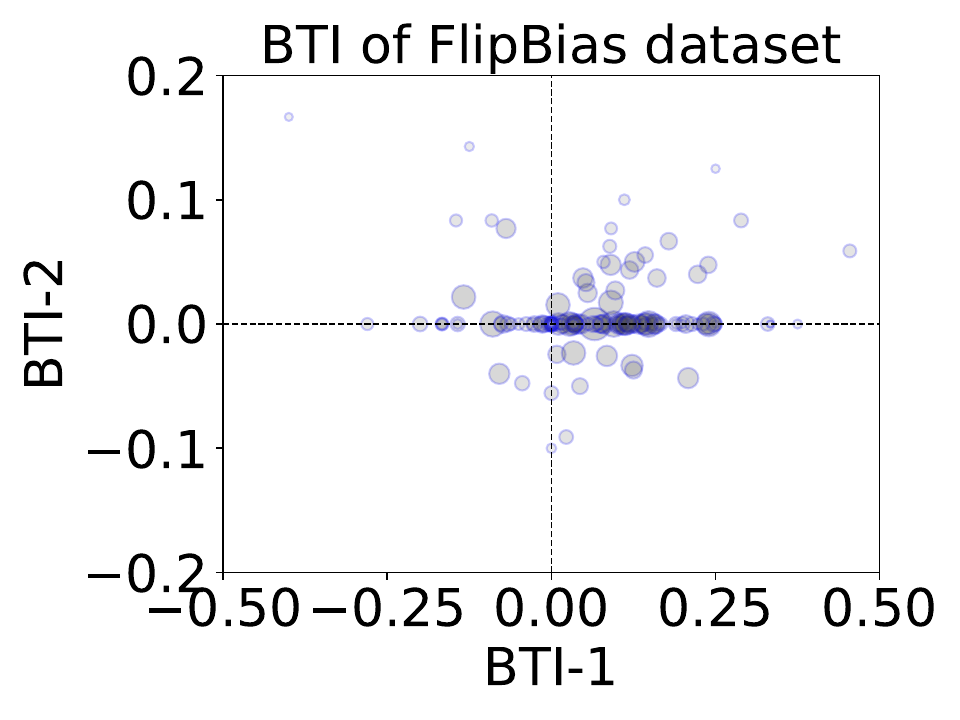}
}
\subfigure[DS] {\label{sfig:FB_debias_topic_v1_v2_DS}
\includegraphics[width=0.225\linewidth]{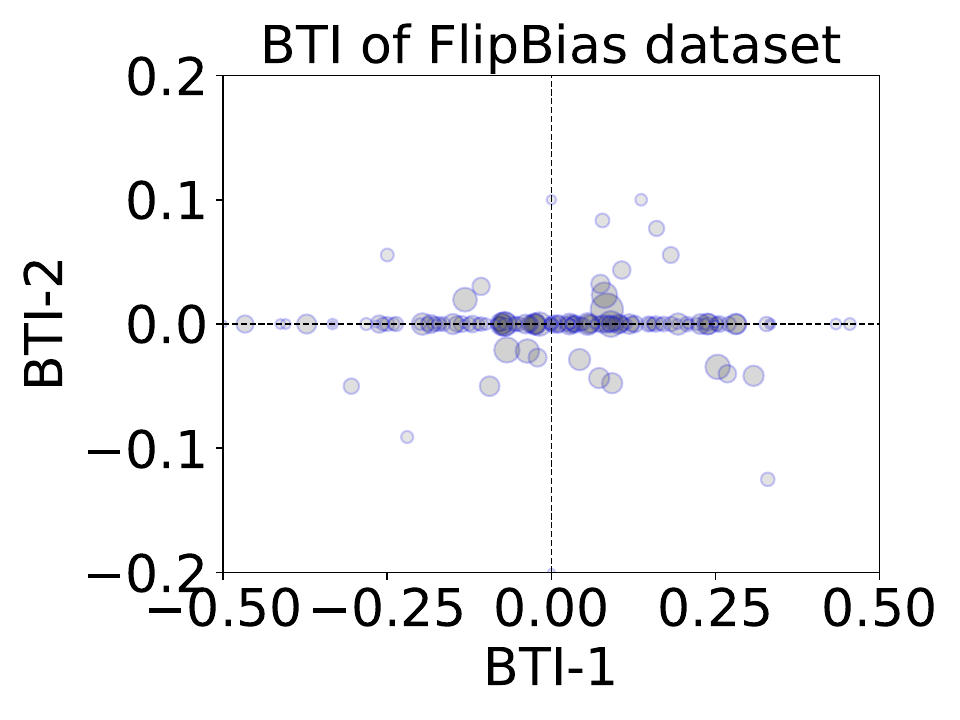}
}
\subfigure[LCR-FT] {\label{sfig:FB_debias_topic_v1_v2_lcrft}
\includegraphics[width=0.225\linewidth]{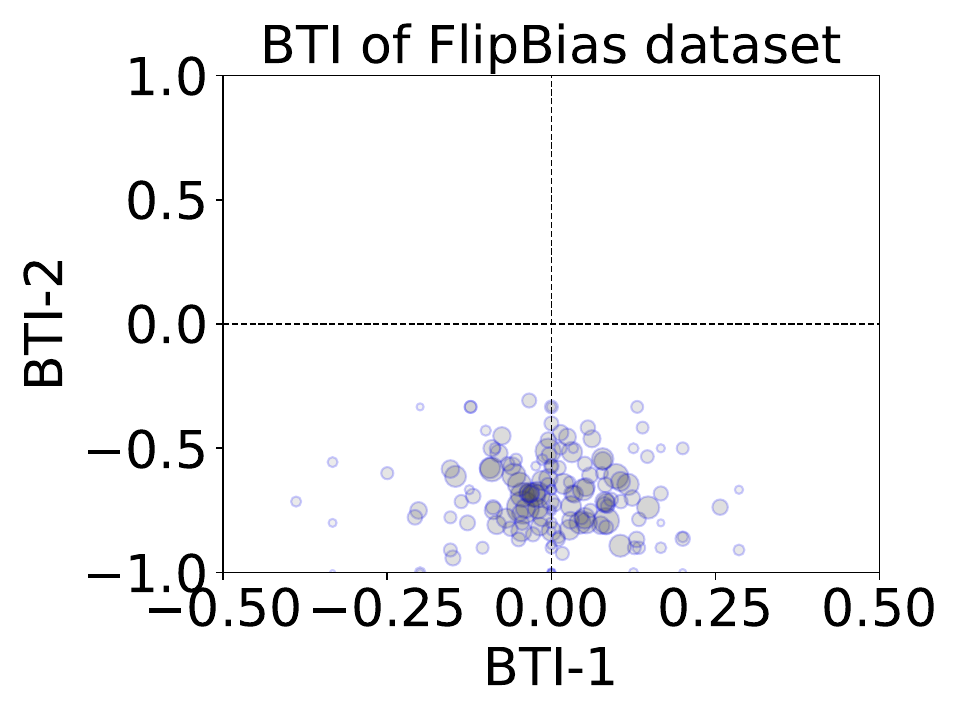}
}
\vskip -1em
\caption{\label{fig:FB_debias_topic_v1_v2} BTI distribution of Topics on FlipBias dataset after debiasing. More distributions are in \Cref{sec:appendix_debias}.
}
\vskip -1em
\end{figure*}
In this section, we delve into effective strategies for debiasing LLMs from two perspectives: strategies in in-context learning and fine-tuning methods for debiasing LLMs. While debiasing exploration is not the primary focus of this work, we present some results of basic debiasing attempts and assess debiasing effectiveness across three aspects: (1) overall debiasing results, (2) elimination of bias across all topics, and (3) potential performance improvements in bias prediction with debiased LLMs.

\subsection{Prompts to Isolate Bias}\label{ssec:prompt_debias}

\paragraph{Bias Label Explanation (BLE).} As we mainly focus on political bias in this work, we incorporate detailed explanations of left-leaning and right-leaning perspectives sourced from Wikipedia to explore if adding bias label explanations to the prompt would help debiasing.

\paragraph{Few-shot Instruction.} A small number (e.g., 3, 6, 12 in our experiment) of examples are explicitly designed to instruct LLMs to recognize biases. Given that FlipBias and ABP operate at the article level, we provide only the headlines to avoid lengthy inputs, as headlines have been proven effective in indicating bias \cite{chen-etal-2018-learning}. Further details of the chosen instances are in \Cref{appendix:subsec:details_prompts_isolate_bias}.

\paragraph{Debiasing Statement (DS).} \citet{ganguli2023capacity} discovered that providing a simple instruction to an LLM, urging it to avoid biases in its responses, can effectively mitigate inherent biases. Following \citet{kaneko2024evaluating}, we add a debiasing statement to our prompts: ``\textit{Please ensure that your answer is unbiased and free from reliance on stereotypes.}''.

\subsection{Fine-Tuning to Debias}\label{ssec:finetune_debias}
By observing the results of \Cref{fig:FB_ABP_ground_predicted}, we infer that the LLM demonstrates better performance in clarifying right-label articles from center-label articles compared to clarifying left-label articles from center-label ones. This observation suggests a potential deficiency in the LLM's ability to accurately recognize left-leaning evaluation criteria. To address this, we adjust the proportion of left-leaning articles in the fine-tuning instances to investigate how varying proportions impact the debiasing process. Specifically, we fine-tune \texttt{gpt-3.5-turbo} using 300 labeled instances (sampled from the regular training sets of datasets)
with three different proportions: all left-label articles (L-FT), a mixture of left-label and center-label articles (LC-FT), and an equal distribution of left-label, center-label, and right-label articles (LCR-FT). We also conduct the LCR-FT setting on the baseline BERT (pre-trained model \texttt{bert-base-cased}) for a better comparison.

\begin{table}[t]
\setlength{\tabcolsep}{1.1mm}\small
\newcommand{\tabincell}[2]{\begin{tabular}{@{}#1@{}}#2\end{tabular}}
\begin{center}
\begin{tabular}{l|cc|ccccc}
\toprule
Models & BTI-1 & BTI-2 & Pre & Rec & BiF1 & MiF1 & MaF1 
\\
\midrule
Vanilla  & 0.06 & 0.01 & 89.0 & 15.4 & 26.2 & 42.3 & 39.4 \\
\midrule
BLE  & 0.03 & 0.00 & 89.3 & 9.4 & 17.0 & 38.8 & 34.3 \\
3-shot & 0.06 & 0.00 & 93.1 & 11.3 & 20.2 & 40.3 & 36.3 \\
6-shot & 0.04 & 0.00 & 92.6 & 9.7 & 17.6 & 39.3 & 34.8 \\
9-shot & 0.04 & 0.00 & \textbf{96.9} & 7.7 & 14.3 & 38.3 & 33.1 \\
DS & 0.01 & 0.00 & 91.9 & 6.7 & 12.4 & 37.4 & 31.8 \\
\midrule
L-FT & 0.00 & -1.00 & 66.7 & \textbf{100.0} & \textbf{80.0} & \textbf{66.7} & 40.0 \\
LC-FT & -0.17 & -0.41 & 67.8 & 43.0 & 52.6 & 48.4 & 48.0 \\
LCR-FT & -0.00 & -0.68 & 68.6 & 89.9 & 77.8 & 65.8 & \textbf{51.7} \\
\midrule
BERT & -0.00 & 0.22 & 66.7 & 99.9 & 79.9 & 66.6 & 39.9 \\

\bottomrule
\end{tabular}
\end{center}
\vskip -1em
\caption{\label{tab:debias_perfromance} 
Debiasing results on FlipBias. 
}
\vskip -1em
\end{table}

\subsection{Assessment of Debiasing Strategies}
We evaluate the debiasing methods introduced in \Cref{ssec:prompt_debias} and \Cref{ssec:finetune_debias} in this subsection. Apart from BTI, the other metrics follow \citet{lin2024indivec}.
\paragraph{General Leaning and Bias Prediction Performance Comparison.} The debiasing results on FlipBias are reported in \Cref{tab:debias_perfromance}. We observe that while finetuning methods generally exhibit better bias prediction performance gains (e.g., better BiF1 and MaF1), they also introduce more bias to the finetuned LLMs, as reflected by larger BTI-1 or BTI-2 values after finetuning. On the other hand, prompt-based debiasing methods show impressive effects, especially DS \cite{ganguli2023capacity}, which is extremely easy yet effective.

\paragraph{Topic-Level Bias Comparison.} We further display the bias tendency index (BTI) distribution on FlipBias after applying some representative debiasing methods in \Cref{fig:FB_debias_topic_v1_v2}, while distributions of additional debiasing methods and results from the ABP dataset can be found in \Cref{sec:appendix_debias}. From \Cref{fig:FB_debias_topic_v1_v2}, we observe that prompt engineering-based debiasing shows better results, as reflected in the BTI values for topics being centered around $0.0$, which is consistent with the general performance comparison results we introduced in the last paragraph. Additionally, the overall shift in the BTI distribution after LCR-FT debiasing, as shown in \Cref{sfig:FB_debias_topic_v1_v2_lcrft}, indicates that finetuning LLMs may result in better performance (refer to bias prediction results reported in \Cref{tab:debias_perfromance}), but it may inadvertently introduce more severe bias.

\paragraph{}
In conclusion, our debiasing experiments reveal that the strategy Bias Label Explanation and Debiasing Statements make the model more cautious in making biased judgments, albeit at the cost of accuracy. In contrast, few-shot and fine-tuning methods yield better performance but are less effective at mitigating bias

\section{RQ4: Do various LLMs exhibit similar bias tendencies?}

\begin{table}[t]
\setlength{\tabcolsep}{1.1mm}\small
\newcommand{\tabincell}[2]{\begin{tabular}{@{}#1@{}}#2\end{tabular}}
\begin{center}
\begin{tabular}{l|cc|ccccc}
\toprule
Models & BTI-1 & BTI-2 & Pre & Rec & BiF1 & MiF1 & MaF1 
\\
\midrule
LLaMa2 & 0.04 & 0.25 & 72.7 & 47.1 & 57.2 & 52.7 & 52.2 \\
Vicuna & -0.01 & 0.07 & 68.0 & 19.1 & 29.8 & 39.8 & 38.5 \\
Mistral & 0.00 & -0.57 & 69.9 & 84.2 & 76.4 & 65.3 & 55.4 \\
\midrule
GPT-3.5  & 0.06 & 0.01 & 89.0 & 15.4 & 26.2 & 42.3 & 39.4 \\
GPT-4 &  0.06 & -0.04 & 85.1 & 30.3 & 44.7 & 50.0 & 49.5  \\

\bottomrule
\end{tabular}
\end{center}
\vskip -1em
\caption{\label{tab:otherllm_perfromance} 
Comparison results of different LLMs. 
}
\vskip -1em
\end{table}
\begin{figure}[t]
\centering
\subfigure[LlaMa2] {\label{sfig:FB_difllms_topic_v1_v2_llama}
\includegraphics[width=0.45\linewidth]{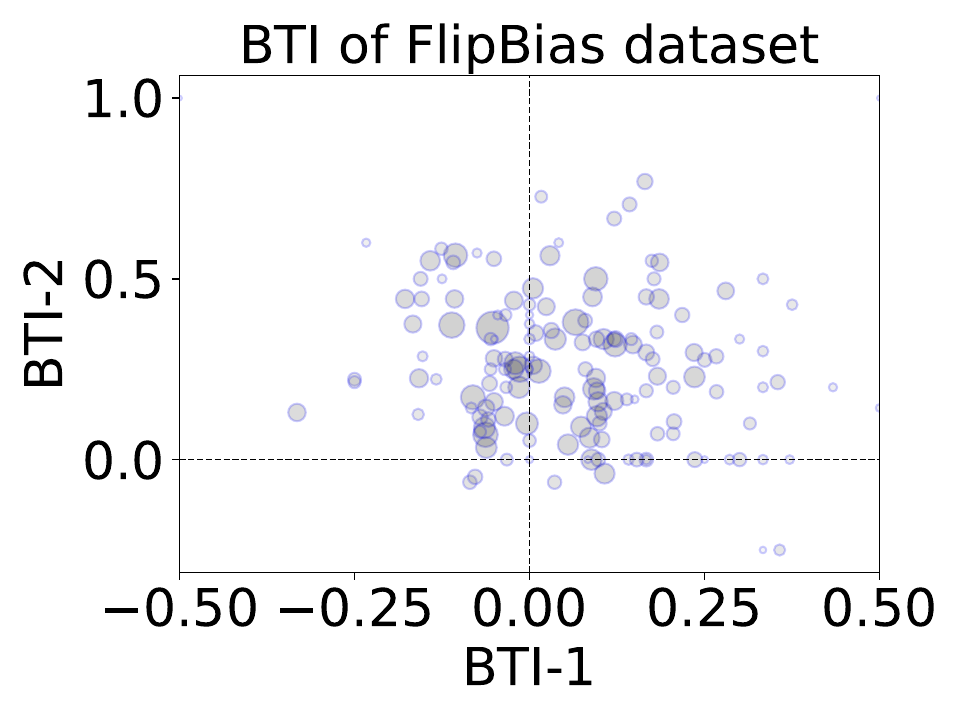}
}
\subfigure[Vicuna] {\label{sfig:FB_difllms_topic_v1_v2_Vicuna}
\includegraphics[width=0.45\linewidth]{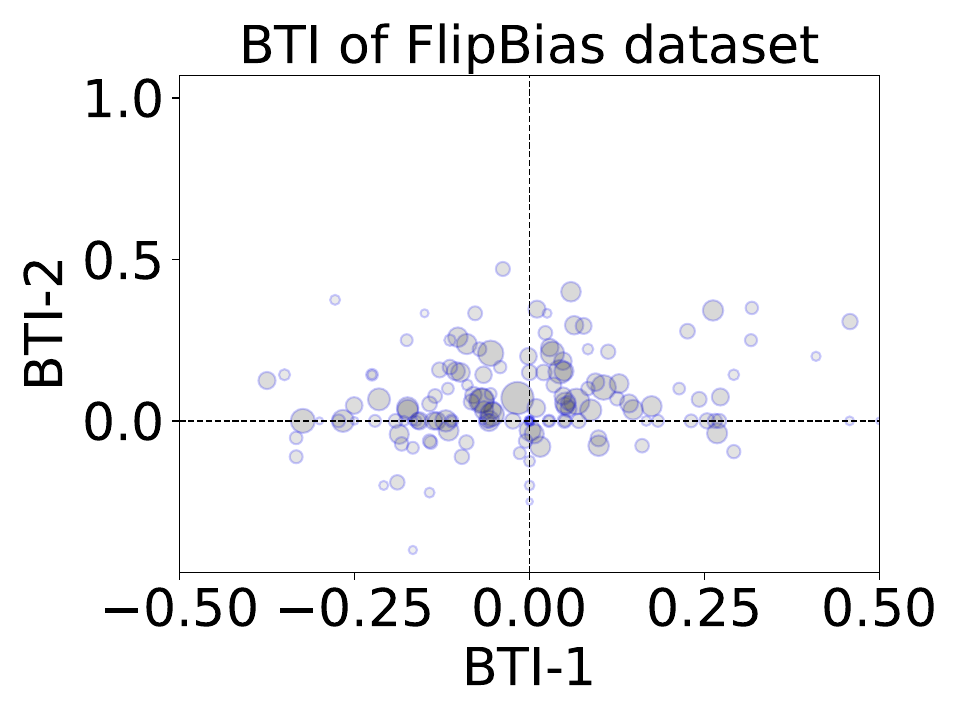}
}
\subfigure[Mistral] {\label{sfig:FB_difllms_topic_v1_v2_Mistral}
\includegraphics[width=0.45\linewidth]{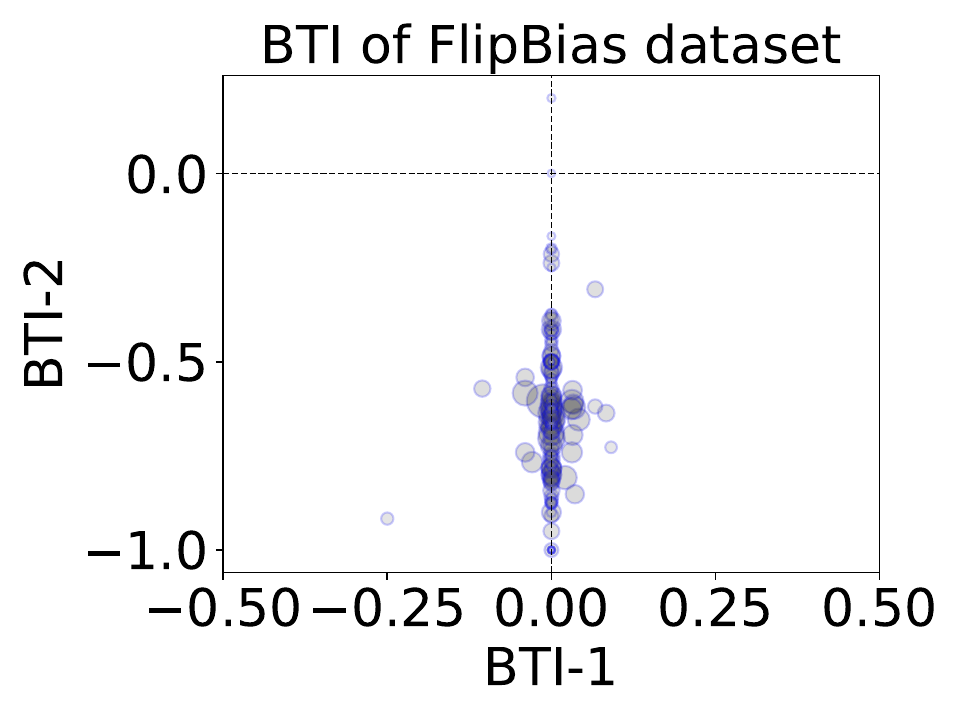}
}
\subfigure[GPT4] {\label{sfig:FB_difllms_topic_v1_v2_gpt4}
\includegraphics[width=0.45\linewidth]{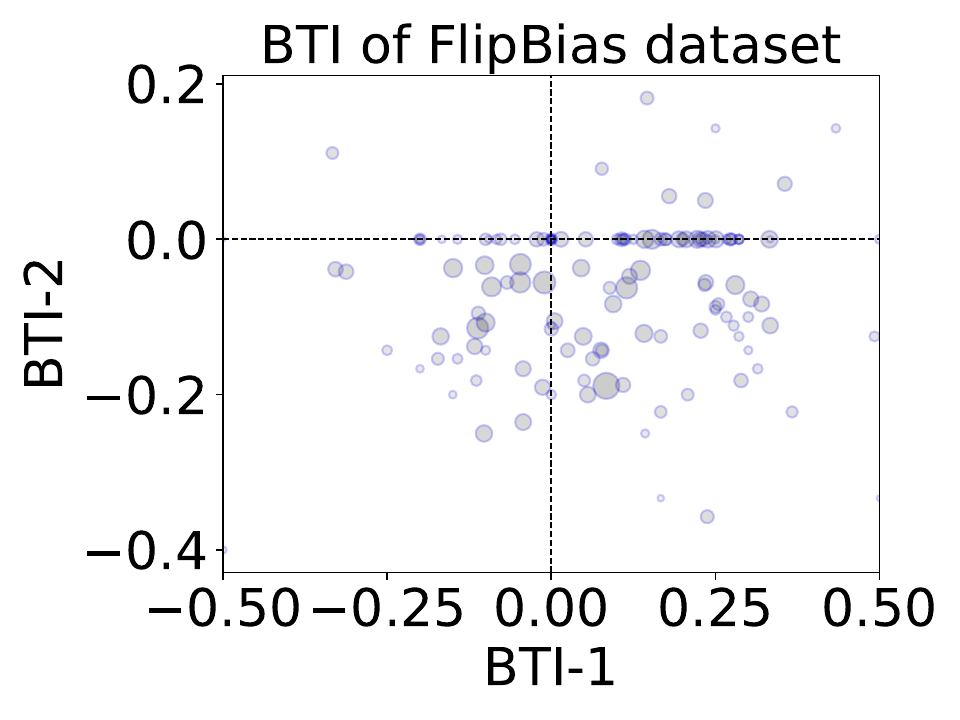}
}
\vskip -1em
\caption{\label{fig:FB_difllms_topic_v1_v2} BTI of Topics on FlipBias for various LLMs.
}
\vskip -1em
\end{figure}
In the previous sections, we conduct experiments using a representative LLM named GPT-3.5. In this section, we extend our analysis to include biases of additional LLMs, both closed-source and open-source. These include \texttt{Llama-2-7B-Chat}, \texttt{Vicuna-7B-v1.5}, \texttt{Mistral-7B-v0.1}, and \texttt{gpt-4-0125-preview}.

We present the bias prediction results and BTI values of these LLMs in \Cref{tab:otherllm_perfromance}, along with the topic-level BTI distribution in \Cref{fig:FB_difllms_topic_v1_v2}. From \Cref{tab:otherllm_perfromance}, it can be observed that LLaMa2 and Mistral even show better political bias performance than GPT-3.5 and GPT-4. However, it is important to clarify that although LLaMa2 and Mistral exhibit better performance according to current classification metrics, they display severe issues such as denying answering and generating unrelated content instead of predicting bias labels (for about 20\% of the testing). Additionally, considering the bias index BTI-1 and BTI-2 values, almost all LLMs exhibit bias, with Mistral showing a general right-leaning tendency, which differs from other LLMs. The fine-grained bias distribution in \Cref{fig:FB_difllms_topic_v1_v2} is consistent with the overall bias reported in \Cref{tab:otherllm_perfromance}.

\section{Conclusion}
In summary, our investigation reveals inherent biases within LLMs and their significant impact on media bias detection. Departing from conventional approaches, we explore biases within LLM systems themselves, particularly in political bias prediction task. Our findings highlight the need for debiasing strategies and provide insights into the broader landscape of bias propagation in language models.

\section*{Limitations}
This work is subject to limitations in two main aspects: (1) Limited Focus on LLM Bias in Media Bias Prediction: The scope of bias analysis is constrained by the availability of three-way (left-, center-, and right-leaning) labeled data. Our study relies on two political bias prediction datasets with three-way labels to investigate biases during LLM prediction. However, datasets with only biased and non-biased labels would not suffice for our analysis in this paper. (2) Assumption of Ground Truth: We operate under the assumption that human-labeled data serves as an unbiased ground truth for assessing LLM biases. Nevertheless, human annotations are inherently subjective and may be influenced by individual biases, potentially impacting the validity of our evaluations.

\section*{Acknowledgments}
This research work is partially supported by CUHK direct grant No. 4055209 and CUHK Knowledge Transfer Project Fund No. KPF23GWP20. We are also grateful to the anonymous reviewers for their comments.

\bibliography{anthology,custom}

\begin{thebibliography}{38}
\expandafter\ifx\csname natexlab\endcsname\relax\def\natexlab#1{#1}\fi

\bibitem[{Baly et~al.(2020)Baly, Martino, Glass, and Nakov}]{baly2020we}
Ramy Baly, Giovanni Da~San Martino, James Glass, and Preslav Nakov. 2020.
\newblock We can detect your bias: Predicting the political ideology of news articles.
\newblock \emph{arXiv preprint arXiv:2010.05338}.

\bibitem[{Bender et~al.(2021)Bender, Gebru, McMillan-Major, and Shmitchell}]{bender2021dangers}
Emily~M Bender, Timnit Gebru, Angelina McMillan-Major, and Shmargaret Shmitchell. 2021.
\newblock On the dangers of stochastic parrots: Can language models be too big?
\newblock In \emph{Proceedings of the 2021 ACM conference on fairness, accountability, and transparency}, pages 610--623.

\bibitem[{Beukeboom and Burgers(2019)}]{beukeboom2019stereotypes}
Camiel~J Beukeboom and Christian Burgers. 2019.
\newblock How stereotypes are shared through language: a review and introduction of the aocial categories and stereotypes communication (scsc) framework.
\newblock \emph{Review of Communication Research}, 7:1--37.

\bibitem[{Blodgett et~al.(2016)Blodgett, Green, and O'Connor}]{blodgett2016demographic}
Su~Lin Blodgett, Lisa Green, and Brendan O'Connor. 2016.
\newblock Demographic dialectal variation in social media: A case study of african-american english.
\newblock \emph{arXiv preprint arXiv:1608.08868}.

\bibitem[{Chen et~al.(2018)Chen, Wachsmuth, Al-Khatib, and Stein}]{chen-etal-2018-learning}
Wei-Fan Chen, Henning Wachsmuth, Khalid Al-Khatib, and Benno Stein. 2018.
\newblock \href {https://doi.org/10.18653/v1/W18-6509} {Learning to flip the bias of news headlines}.
\newblock In \emph{Proceedings of the 11th International Conference on Natural Language Generation}, pages 79--88, Tilburg University, The Netherlands. Association for Computational Linguistics.

\bibitem[{Conti and Wisniewski(2023)}]{conti-wisniewski-2023-using}
Lina Conti and Guillaume Wisniewski. 2023.
\newblock \href {https://doi.org/10.18653/v1/2023.emnlp-main.641} {Using artificial {F}rench data to understand the emergence of gender bias in transformer language models}.
\newblock In \emph{Proceedings of the 2023 Conference on Empirical Methods in Natural Language Processing}, pages 10362--10371, Singapore. Association for Computational Linguistics.

\bibitem[{Esiobu et~al.(2023)Esiobu, Tan, Hosseini, Ung, Zhang, Fernandes, Dwivedi-Yu, Presani, Williams, and Smith}]{esiobu2023robbie}
David Esiobu, Xiaoqing Tan, Saghar Hosseini, Megan Ung, Yuchen Zhang, Jude Fernandes, Jane Dwivedi-Yu, Eleonora Presani, Adina Williams, and Eric Smith. 2023.
\newblock Robbie: Robust bias evaluation of large generative language models.
\newblock In \emph{Proceedings of the 2023 Conference on Empirical Methods in Natural Language Processing}, pages 3764--3814.

\bibitem[{Fang et~al.(2023)Fang, Che, Mao, Zhang, Zhao, and Zhao}]{fang2023bias}
Xiao Fang, Shangkun Che, Minjia Mao, Hongzhe Zhang, Ming Zhao, and Xiaohang Zhao. 2023.
\newblock Bias of ai-generated content: an examination of news produced by large language models.
\newblock \emph{arXiv preprint arXiv:2309.09825}.

\bibitem[{Gallegos et~al.(2023)Gallegos, Rossi, Barrow, Tanjim, Kim, Dernoncourt, Yu, Zhang, and Ahmed}]{gallegos2023bias}
Isabel~O Gallegos, Ryan~A Rossi, Joe Barrow, Md~Mehrab Tanjim, Sungchul Kim, Franck Dernoncourt, Tong Yu, Ruiyi Zhang, and Nesreen~K Ahmed. 2023.
\newblock Bias and fairness in large language models: A survey.
\newblock \emph{arXiv preprint arXiv:2309.00770}.

\bibitem[{Ganguli et~al.(2023)Ganguli, Askell, Schiefer, Liao, Luko{\v{s}}i{\=u}t{\.e}, Chen, Goldie, Mirhoseini, Olsson, Hernandez et~al.}]{ganguli2023capacity}
Deep Ganguli, Amanda Askell, Nicholas Schiefer, Thomas Liao, Kamil{\.e} Luko{\v{s}}i{\=u}t{\.e}, Anna Chen, Anna Goldie, Azalia Mirhoseini, Catherine Olsson, Danny Hernandez, et~al. 2023.
\newblock The capacity for moral self-correction in large language models.
\newblock \emph{arXiv preprint arXiv:2302.07459}.

\bibitem[{Garimella et~al.(2022)Garimella, Mihalcea, and Amarnath}]{garimella2022demographic}
Aparna Garimella, Rada Mihalcea, and Akhash Amarnath. 2022.
\newblock Demographic-aware language model fine-tuning as a bias mitigation technique.
\newblock In \emph{Proceedings of the 2nd Conference of the Asia-Pacific Chapter of the Association for Computational Linguistics and the 12th International Joint Conference on Natural Language Processing}, pages 311--319.

\bibitem[{Gira et~al.(2022)Gira, Zhang, and Lee}]{gira2022debiasing}
Michael Gira, Ruisu Zhang, and Kangwook Lee. 2022.
\newblock Debiasing pre-trained language models via efficient fine-tuning.
\newblock In \emph{Proceedings of the Second Workshop on Language Technology for Equality, Diversity and Inclusion}, pages 59--69.

\bibitem[{Gon{\c{c}}alves and Strubell(2023)}]{gonccalves2023understanding}
Gustavo Gon{\c{c}}alves and Emma Strubell. 2023.
\newblock Understanding the effect of model compression on social bias in large language models.
\newblock In \emph{Proceedings of the 2023 Conference on Empirical Methods in Natural Language Processing}, pages 2663--2675.

\bibitem[{Hada et~al.(2023)Hada, Seth, Diddee, and Bali}]{hada2023fifty}
Rishav Hada, Agrima Seth, Harshita Diddee, and Kalika Bali. 2023.
\newblock “fifty shades of bias”: Normative ratings of gender bias in gpt generated english text.
\newblock In \emph{Proceedings of the 2023 Conference on Empirical Methods in Natural Language Processing}, pages 1862--1876.

\bibitem[{Iyyer et~al.(2014)Iyyer, Enns, Boyd-Graber, and Resnik}]{iyyer2014political}
Mohit Iyyer, Peter Enns, Jordan Boyd-Graber, and Philip Resnik. 2014.
\newblock Political ideology detection using recursive neural networks.
\newblock In \emph{Proceedings of the 52nd Annual Meeting of the Association for Computational Linguistics (Volume 1: Long Papers)}, pages 1113--1122.

\bibitem[{Joniak and Aizawa(2022)}]{joniak2022gender}
Przemyslaw Joniak and Akiko Aizawa. 2022.
\newblock Gender biases and where to find them: Exploring gender bias in pre-trained transformer-based language models using movement pruning.
\newblock \emph{arXiv preprint arXiv:2207.02463}.

\bibitem[{Kaneko et~al.(2024)Kaneko, Bollegala, Okazaki, and Baldwin}]{kaneko2024evaluating}
Masahiro Kaneko, Danushka Bollegala, Naoaki Okazaki, and Timothy Baldwin. 2024.
\newblock Evaluating gender bias in large language models via chain-of-thought prompting.
\newblock \emph{arXiv preprint arXiv:2401.15585}.

\bibitem[{Lin et~al.()Lin, Wang, Guo, Li, and Wong}]{lin2024inditag}
Luyang Lin, Lingzhi Wang, Jinsong Guo, Jing Li, and Kam-Fai Wong.
\newblock \href {http://arxiv.org/abs/2403.13446} {Inditag: An online media bias analysis and annotation system using fine-grained bias indicators}.

\bibitem[{Lin et~al.(2024)Lin, Wang, Zhao, Li, and Wong}]{lin2024indivec}
Luyang Lin, Lingzhi Wang, Xiaoyan Zhao, Jing Li, and Kam-Fai Wong. 2024.
\newblock Indivec: An exploration of leveraging large language models for media bias detection with fine-grained bias indicators.
\newblock \emph{arXiv preprint arXiv:2402.00345}.

\bibitem[{Liu et~al.(2024)Liu, Wang, Li, and Li}]{liu2024teller}
Hui Liu, Wenya Wang, Haoru Li, and Haoliang Li. 2024.
\newblock Teller: A trustworthy framework for explainable, generalizable and controllable fake news detection.
\newblock \emph{arXiv preprint arXiv:2402.07776}.

\bibitem[{Liu et~al.(2021)Liu, Jia, Wei, Xu, Wang, and Vosoughi}]{liu2021mitigating}
Ruibo Liu, Chenyan Jia, Jason Wei, Guangxuan Xu, Lili Wang, and Soroush Vosoughi. 2021.
\newblock Mitigating political bias in language models through reinforced calibration.
\newblock In \emph{Proceedings of the AAAI Conference on Artificial Intelligence}, volume~35, pages 14857--14866.

\bibitem[{Liu et~al.(2022)Liu, Zhang, Wegsman, Beauchamp, and Wang}]{liu2022politics}
Yujian Liu, Xinliang~Frederick Zhang, David Wegsman, Nick Beauchamp, and Lu~Wang. 2022.
\newblock Politics: pretraining with same-story article comparison for ideology prediction and stance detection.
\newblock \emph{arXiv preprint arXiv:2205.00619}.

\bibitem[{Motoki et~al.(2024)Motoki, Pinho~Neto, and Rodrigues}]{motoki2024more}
Fabio Motoki, Valdemar Pinho~Neto, and Victor Rodrigues. 2024.
\newblock More human than human: Measuring chatgpt political bias.
\newblock \emph{Public Choice}, 198(1):3--23.

\bibitem[{Peng et~al.(2023)Peng, Liu, Yang, Yuan, and Li}]{peng2023embedding}
Ruoling Peng, Kang Liu, Po~Yang, Zhipeng Yuan, and Shunbao Li. 2023.
\newblock Embedding-based retrieval with llm for effective agriculture information extracting from unstructured data.
\newblock \emph{arXiv preprint arXiv:2308.03107}.

\bibitem[{Qian et~al.(2022)Qian, Ross, Fernandes, Smith, Kiela, and Williams}]{qian2022perturbation}
Rebecca Qian, Candace Ross, Jude Fernandes, Eric Smith, Douwe Kiela, and Adina Williams. 2022.
\newblock Perturbation augmentation for fairer nlp.
\newblock \emph{arXiv preprint arXiv:2205.12586}.

\bibitem[{Rozado(2023)}]{rozado2023political}
David Rozado. 2023.
\newblock The political biases of chatgpt.
\newblock \emph{Social Sciences}, 12(3):148.

\bibitem[{Rozado(2024)}]{rozado2024political}
David Rozado. 2024.
\newblock The political preferences of llms.
\newblock \emph{arXiv preprint arXiv:2402.01789}.

\bibitem[{Smith et~al.(2022)Smith, Hall, Kambadur, Presani, and Williams}]{smith2022m}
Eric~Michael Smith, Melissa Hall, Melanie Kambadur, Eleonora Presani, and Adina Williams. 2022.
\newblock “i’m sorry to hear that”: Finding new biases in language models with a holistic descriptor dataset.
\newblock In \emph{Proceedings of the 2022 Conference on Empirical Methods in Natural Language Processing}, pages 9180--9211.

\bibitem[{Taubenfeld et~al.(2024)Taubenfeld, Dover, Reichart, and Goldstein}]{taubenfeld2024systematic}
Amir Taubenfeld, Yaniv Dover, Roi Reichart, and Ariel Goldstein. 2024.
\newblock Systematic biases in llm simulations of debates.
\newblock \emph{arXiv preprint arXiv:2402.04049}.

\bibitem[{Tokpo and Calders(2022)}]{tokpo2022text}
Ewoenam~Kwaku Tokpo and Toon Calders. 2022.
\newblock Text style transfer for bias mitigation using masked language modeling.
\newblock \emph{arXiv preprint arXiv:2201.08643}.

\bibitem[{Urman and Makhortykh(2023)}]{urman2023silence}
Aleksandra Urman and Mykola Makhortykh. 2023.
\newblock The silence of the llms: Cross-lingual analysis of political bias and false information prevalence in chatgpt, google bard, and bing chat.

\bibitem[{Venkit et~al.(2023)Venkit, Gautam, Panchanadikar, Huang, and Wilson}]{venkit2023nationality}
Pranav~Narayanan Venkit, Sanjana Gautam, Ruchi Panchanadikar, Ting-Hao'Kenneth' Huang, and Shomir Wilson. 2023.
\newblock Nationality bias in text generation.
\newblock \emph{arXiv preprint arXiv:2302.02463}.

\bibitem[{Wambsganss et~al.(2023{\natexlab{a}})Wambsganss, Su, Swamy, Neshaei, Rietsche, and K{\"a}ser}]{wambsganss-etal-2023-unraveling}
Thiemo Wambsganss, Xiaotian Su, Vinitra Swamy, Seyed Neshaei, Roman Rietsche, and Tanja K{\"a}ser. 2023{\natexlab{a}}.
\newblock \href {https://doi.org/10.18653/v1/2023.findings-emnlp.689} {Unraveling downstream gender bias from large language models: A study on {AI} educational writing assistance}.
\newblock In \emph{Findings of the Association for Computational Linguistics: EMNLP 2023}, pages 10275--10288, Singapore. Association for Computational Linguistics.

\bibitem[{Wambsganss et~al.(2023{\natexlab{b}})Wambsganss, Su, Swamy, Neshaei, Rietsche, and K{\"a}ser}]{wambsganss2023unraveling}
Thiemo Wambsganss, Xiaotian Su, Vinitra Swamy, Seyed~Parsa Neshaei, Roman Rietsche, and Tanja K{\"a}ser. 2023{\natexlab{b}}.
\newblock Unraveling downstream gender bias from large language models: A study on ai educational writing assistance.
\newblock \emph{arXiv preprint arXiv:2311.03311}.

\bibitem[{Wang et~al.(2023)Wang, Mo, Wang, Zhou, and Chen}]{wang2023causal}
Fei Wang, Wenjie Mo, Yiwei Wang, Wenxuan Zhou, and Muhao Chen. 2023.
\newblock A causal view of entity bias in (large) language models.
\newblock \emph{arXiv preprint arXiv:2305.14695}.

\bibitem[{Yano et~al.(2010)Yano, Resnik, and Smith}]{yano2010shedding}
Tae Yano, Philip Resnik, and Noah~A Smith. 2010.
\newblock Shedding (a thousand points of) light on biased language.
\newblock In \emph{Proceedings of the NAACL HLT 2010 Workshop on Creating Speech and Language Data with Amazon’s Mechanical Turk}, pages 152--158.

\bibitem[{Yu et~al.(2008)Yu, Kaufmann, and Diermeier}]{yu2008classifying}
Bei Yu, Stefan Kaufmann, and Daniel Diermeier. 2008.
\newblock Classifying party affiliation from political speech.
\newblock \emph{Journal of Information Technology \& Politics}, 5(1):33--48.

\bibitem[{Zayed et~al.(2023)Zayed, Parthasarathi, Mordido, Palangi, Shabanian, and Chandar}]{zayed2023deep}
Abdelrahman Zayed, Prasanna Parthasarathi, Gon{\c{c}}alo Mordido, Hamid Palangi, Samira Shabanian, and Sarath Chandar. 2023.
\newblock Deep learning on a healthy data diet: Finding important examples for fairness.
\newblock In \emph{Proceedings of the AAAI Conference on Artificial Intelligence}, volume~37, pages 14593--14601.

\end{thebibliography}
\newpage
\appendix
\section{Left-Right Vocabulary Corpus Construction}
\label{appendix:sec:left_right_vocabulary_construction}
We construct the Left-Right vocabulary corpus using the ABP dataset. Initially, all articles in ABP are tokenized using the NLTK Python package. Tokens are converted to lowercase and filtered using a stopwords corpus. Each token is then labeled based on the articles they appear in.

To create the Left-Right Vocabulary Corpus, we prioritize tokens labeled with significantly higher frequencies in either Left or Right articles. Specifically, we calculate the Left ratio by dividing a token's frequency in Left articles by the total tokens in Left articles, and similarly for the Right ratio. Tokens are included in the Left vocabulary list only if the Left ratio is more than twice the Right ratio.

From the Left vocabulary list, we select the top 2000 most frequent tokens. We then select 1295 tokens from the Right vocabulary list to match the total frequency sum of the Left tokens. This corpus is validated against the vocabulary of \citet{yano2010shedding}. The constructed vocabularies will be publicly available for future research.

\begin{table*}[t]
\setlength{\tabcolsep}{1mm}\small
\begin{center}
\begin{tabular}{p{12cm}p{2.5cm}}
\toprule[1.0pt]
Indicators & Interpreted Topic
\\
\midrule
"Provides figures and quotes from individuals involved in the issue", "The article cites statements from various individuals involved in the case, including lawyers, politicians, and advocacy groups.", "The text quotes various experts and government officials to support its claims.", "cites tweets and quotes from Trump, experts, and state officials to support the claims made", "Quotes from various food experts and diplomats.", "The article cites multiple sources, including government documents and quotes from officials." ... ...  & Comprehensive Use of Quotes and Citations in Journalism \\
\midrule
"Describes President Obama's decision as "benighted" and "cowardly" while praising President Trump's decision", ""swipes at Joe Biden," "knocks primary rival Bernie Sanders," "gripes about former President Barack Obama"", "frames the issue as a result of understaffing and mismanagement, blames the Obama administration, and highlights the need to protect the president", "Celebratory tone towards Obama, sarcastic and mocking tone towards Democrats", "Portrays Democrats as wanting a grander celebration, mocks Obama and the holiday", "Frames the decision as a potential unwinding of an Obama executive action, includes criticism from Democrats and environmental groups", "Describes the tough-on-crime approach as a reversal of Obama's "Smart on Crime" policy, implying a negative change" ... ...  & Diverse Perspectives on President Obama’s Policies and Actions\\
\midrule
"Mentions celebrations and security measures in various cities", "Mentions specific incidents of terrorism and security measures in different cities", "Presents the incident as a terrorist attack and highlights the victims' nationalities", "Mentions previous vehicle attacks and quotes from witnesses", "Provides examples of other major music event attacks", "Mentions the arsenal of weapons and ammunition recovered, suggesting the possibility of an accomplice", "Mentions the London subway station fire as a terrorist incident caused by an improvised explosive device", "The article provides examples of previous attacks and the use of improvised explosive devices." ... ... & Analysis of Recent Terrorist Attacks and Security Measures in Various Cities\\
\bottomrule
\end{tabular}
\end{center}
\caption{\label{app:tab:clustered_indicator} 
Clustered Indicators and Interpreted Topics.
}
\end{table*}

\begin{table*}[t]
\setlength{\tabcolsep}{1mm}\small
\begin{center}
\begin{tabular}{p{12cm}|ccc}
\toprule[1.0pt]
Interpretation of Top and Bottom 5 latent
topics (ranked by BTI-2 values)  &BTI-1& BTI-2& Frequency  \\
\midrule[0.5pt]
Trump's Clashes with Federal Law Enforcement and Media &-0.17&\cellcolor{blue!11}0.11& 80 \\
Examining Controversial Tactics: Dissecting Allegations and Defenses in Recent Political Affairs &0.18&\cellcolor{blue!10}0.10& 64 \\
Analysis of Congressional Dynamics: Trump's Strategy, Witness Battles, and Financial Focus & -0.05 & \cellcolor{blue!9}0.09 & 72 \\
Bipartisan Cooperation in Senate: Struggles and Progress & -0.05 & \cellcolor{blue!8}0.08 &  68 \\
Unveiling the Constitutional Crisis: Examining Government Overreach and the Erosion of Rights & 0.25 & \cellcolor{blue!7} 0.07 & 72\\
\midrule
Understanding Textual Analysis: The Importance of Examples and Analogies & 0.18 & \cellcolor{orange!5}-0.05 & 59 \\
Statewide Controversies: Voter Rights, Criminal Justice, and Transition Integrity & 0.18 & \cellcolor{orange!6}-0.06 & 60 \\
Analyzing Political Discourse: Insights from Trump Administration and Beyond &  -0.02 & \cellcolor{orange!8}-0.08 & 76 \\
Examining Biased Reporting in Political Discourse: Imbalance in State of the Union Addresses & 0.20 & \cellcolor{orange!9}-0.09 &  96 \\
Critical Discourse Analysis of Media Portrayal on Trump's Governance &-0.01 & \cellcolor{orange!10}-0.10 & 91 \\
\bottomrule[1.0pt]
\end{tabular}
\end{center}
\caption{\label{app:tab:appendix_fb_topic_case_study} Interpretation of Top and Bottom 5 Latent Topics on FlipBias.}
\end{table*}

\section{Latent Topic Construction}
\label{appendix:sec:latent_topic_construction}

Inspired by IndiVec \cite{lin2024indivec}, we prompted ChatGPT to construct fine-grained media bias indicators using the Flipbias dataset. These indicators summarize key points that may reflect media bias in each article. To organize the topics covered in these articles more effectively, we performed strict clustering through Hierarchical Clustering based on Euclidean distance applied to the indicators extracted from Flipbias. We utilized AgglomerativeClustering from the Scikit Learn package, setting the distance threshold to 2. The embeddings of the indicators were derived from OpenAIEmbeddings. Ultimately, 19,671 indicators were clustered into 152 clusters, each representing a latent topic.

\paragraph{Latent Topics and Corresponding Clustered Indicators}
Details of the clustered indicators are provided in \Cref{app:tab:clustered_indicator}.

\paragraph{Latent Topic Cases Ranked by BTI-2 Values}
Rankings of latent topic cases based on BTI-2 values are shown in \Cref{app:tab:appendix_fb_topic_case_study}.

\section{Implementation Details}
\subsection{Details of Prompts to Isolate Bias}
\label{appendix:subsec:details_prompts_isolate_bias}
\begin{table*}[t]
\setlength{\tabcolsep}{1mm}\small
\begin{center}
\begin{tabular}{p{14cm}}
\toprule[1.0pt]
\textbf{Left}-wing politics describes the range of political ideologies that support and seek to achieve social equality and egalitarianism, often in opposition to social hierarchy as a whole or certain social hierarchies. Left-wing politics typically involve a concern for those in society whom its adherents perceive as disadvantaged relative to others as well as a belief that there are unjustified inequalities that need to be reduced or abolished through radical means that change the nature of the society they are implemented in. \\
\midrule
\textbf{Right}-wing politics is the range of political ideologies that view certain social orders and hierarchies as inevitable, natural, normal, or desirable, typically supporting this position based on natural law, economics, authority, property or tradition. Hierarchy and inequality may be seen as natural results of traditional social differences or competition in market economies.\\
\midrule
\textbf{Centrism} is a political outlook or position involving acceptance or support of a balance of social equality and a degree of social hierarchy while opposing political changes that would result in a significant shift of society strongly to the left or the right. \\
\bottomrule[1.0pt]
\end{tabular}
\end{center}
\caption{Examples of Article Continuation}
\label{tab:appendix_explanation_ble}
\end{table*}

In \Cref{ssec:prompt_debias}, we discussed the methods to debias LLMs. Here we provie details of these debaising methods. 
\paragraph{Bias Label Explanation} In Bias Label Explanation (BLE) method, we adopt explanations as listed in \Cref{tab:appendix_explanation_ble}.

\paragraph{Few-shot Instances} In Few-shot Instruction method, we randomly selected 4 Left-Center-Right triples from the dataset Flipbias and then used the titles as the instances of few-shot instruction, which are listed in \Cref{app:tab:fewshot_instances}.

\begin{table*}[t]
\setlength{\tabcolsep}{1mm}\small
\begin{center}
\begin{tabular}{p{14cm}l}
\toprule[1.0pt]
Text & Label 
\\
\midrule
Trump Accuses His Justice Department, FBI Of Favoring Democrats & Left  \\
Explosive memo released as Trump escalates fight over Russia probe & Center \\
Trump accuses FBI, DOJ leadership of bias against Republicans and in favor of Dems & Right\\
\midrule
Shutdown truce just delays Trump's big dilemma & Left \\
Winners and losers from the government shutdown& Center\\
Centrists break Senate logjam, pave new path for ‘common sense’ bipartisanship  & Right\\
\midrule
North Korean insults to U.S. leaders are nothing new — but Trump’s deeply personal reactions are & Left \\
Trump trades 'short and fat' barb with N Korea's Kim & Center\\
Trump Take To Social Media To Hit Back At 'Short and Fat' Kim Jong-un  & Right\\
\midrule
After 16 Futile Years, Congress Will Try Again to Legalize ‘Dreamers’ & Left \\
The clock is ticking': Graham and Durbin urge action on bipartisan DREAM Act by the end of September & Center\\
Republican Sen. Cory Gardner agrees to support bipartisan Dream Act after Trump rescinds DACA  & Right \\
\bottomrule
\end{tabular}
\end{center}
\caption{\label{app:tab:fewshot_instances} 
few-shot instances
}
\end{table*}

\subsection{More Finetuning Implementation Details}
\label{appendix:sec:finetuning_implementation_details}
\paragraph{GPT Finetuning Details}
We fine-tuned gpt-3.5-turbo through the API supplied by OpenAI. 300 instances are randomly selected from the dataset ABP according to our setting as the training set. The hyperparameters of the number of epochs is 3 and the batch size is 32.

\section{Debiasing Results}\label{sec:appendix_debias}
\begin{figure*}[ht]
\centering
\subfigure[BLE] {\label{sfig:ABP_debias_topic_v1_v2_BLE}
\includegraphics[width=0.25\linewidth]{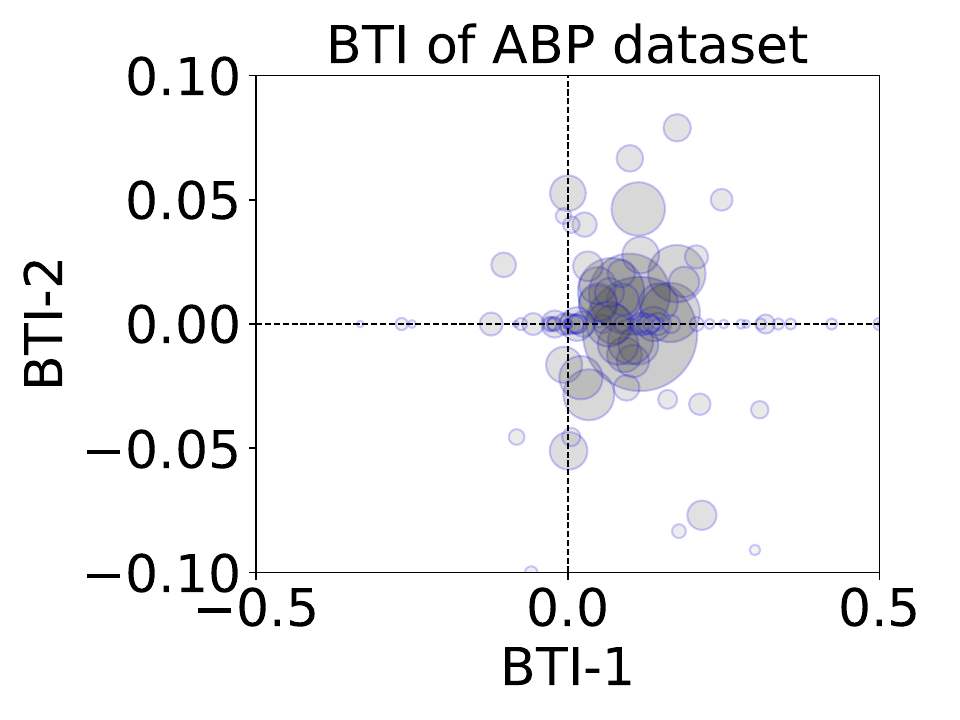}
}
\subfigure[3-Shot] {\label{sfig:ABP_debias_topic_v1_v2_3shot}
\includegraphics[width=0.25\linewidth]{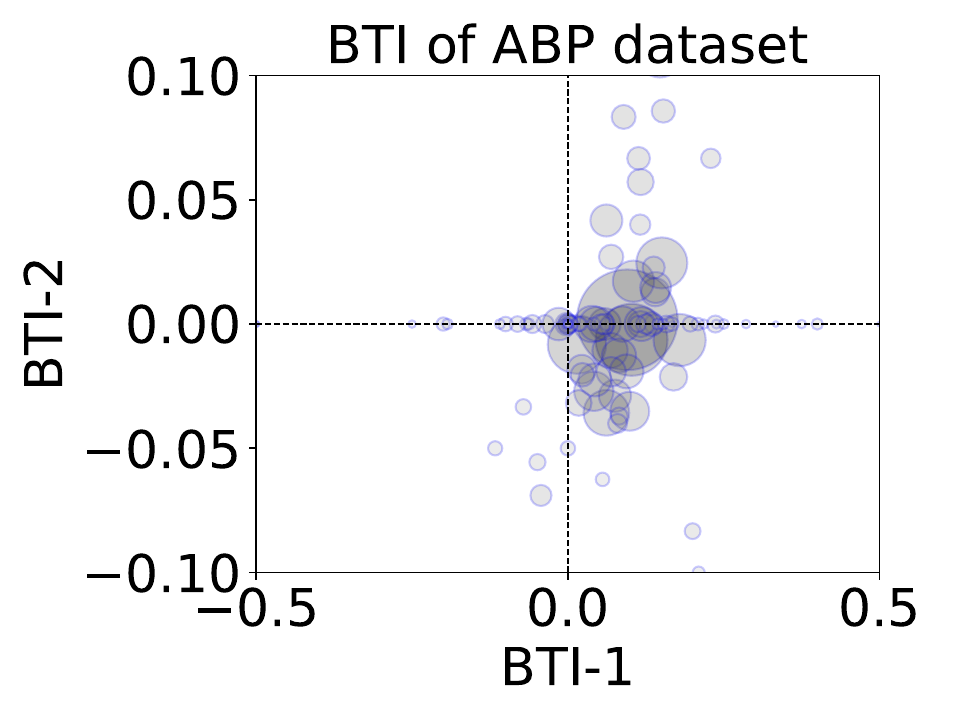}
}
\subfigure[6-Shot] {\label{sfig:ABP_debias_topic_v1_v2_6shot}
\includegraphics[width=0.25\linewidth]{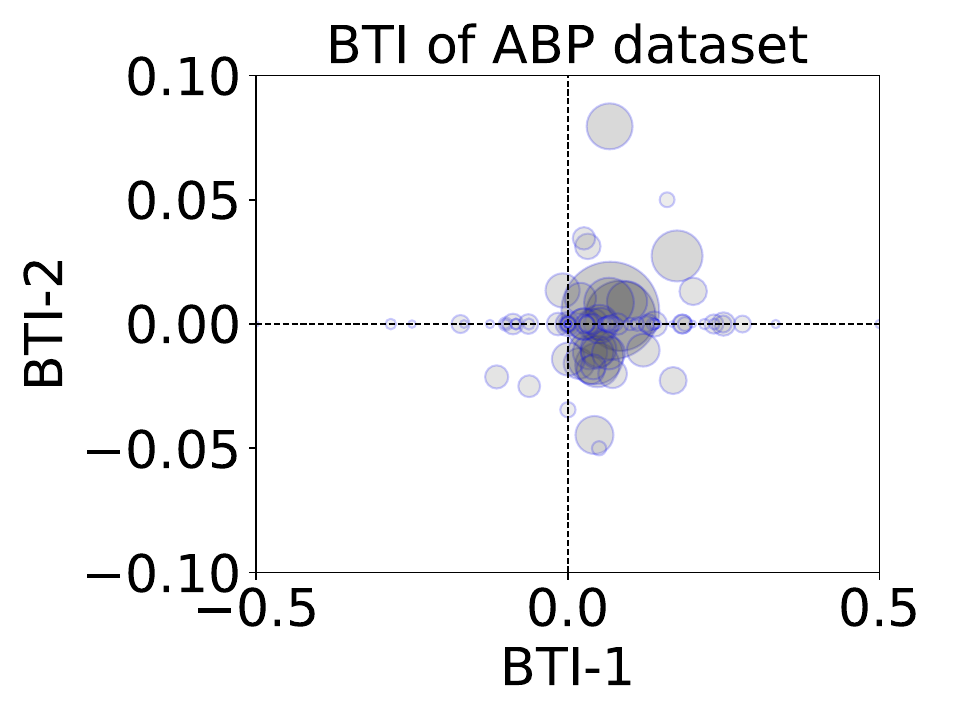}
}
\subfigure[9-Shot] {\label{sfig:ABP_debias_topic_v1_v2_9shot}
\includegraphics[width=0.25\linewidth]{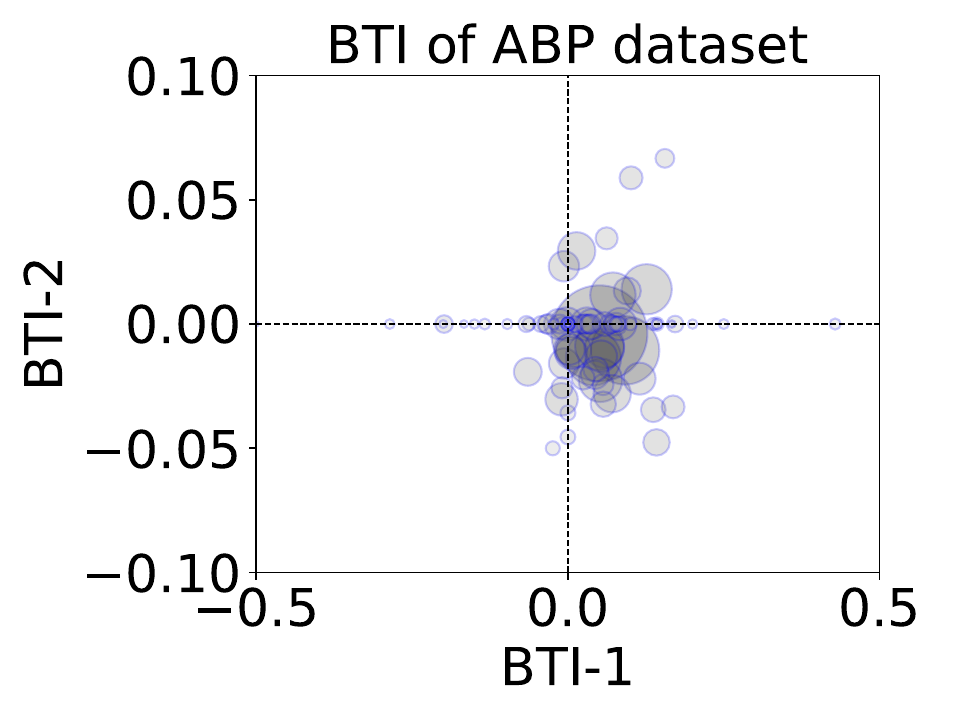}
}
\subfigure[DS] {\label{sfig:ABP_debias_topic_v1_v2_DS}
\includegraphics[width=0.25\linewidth]{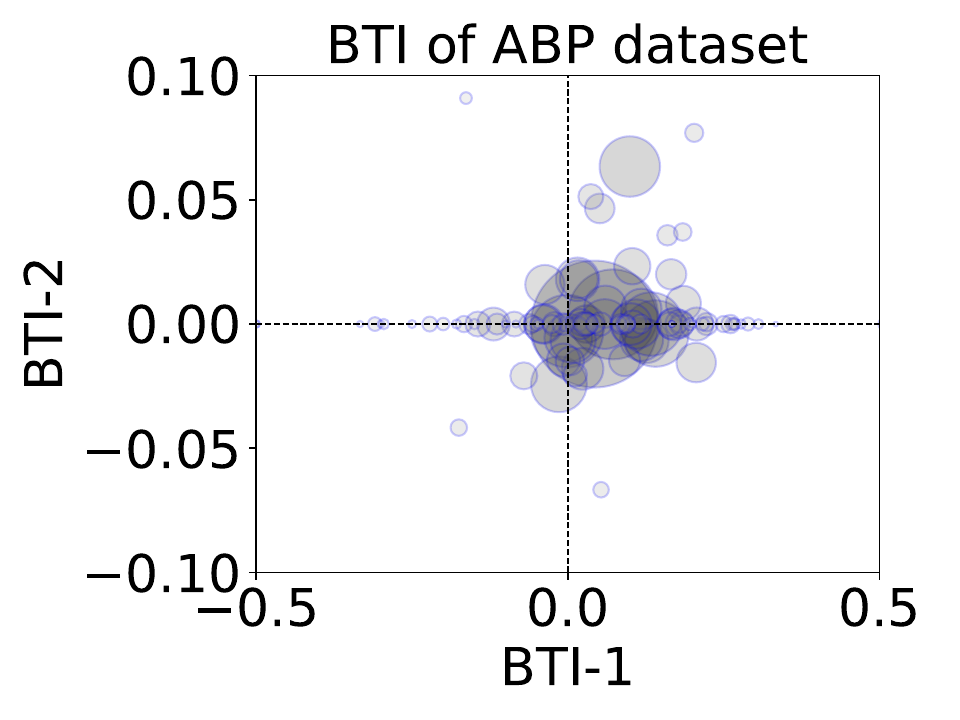}
}
\vskip -1em
\caption{\label{fig:ABP_debias_topic_v1_v2} BTI distribution of Topics on ABP dataset after prompt debiasing.
}
\vskip -1em
\end{figure*}
\begin{figure*}[ht]
\centering
\subfigure[BLE] {\label{sfig:appendix_FB_debias_topic_v1_v2_BLE}
\includegraphics[width=0.25\linewidth]{latex/pictures/debias_topic_FB_v1-v2_ble.pdf}
}
\subfigure[3-Shot] {\label{sfig:appendix_FB_debias_topic_v1_v2_3shot}
\includegraphics[width=0.25\linewidth]{latex/pictures/debias_topic_FB_v1-v2_3shot.pdf}
}
\subfigure[6-Shot] {\label{sfig:appendix_FB_debias_topic_v1_v2_6shot}
\includegraphics[width=0.25\linewidth]{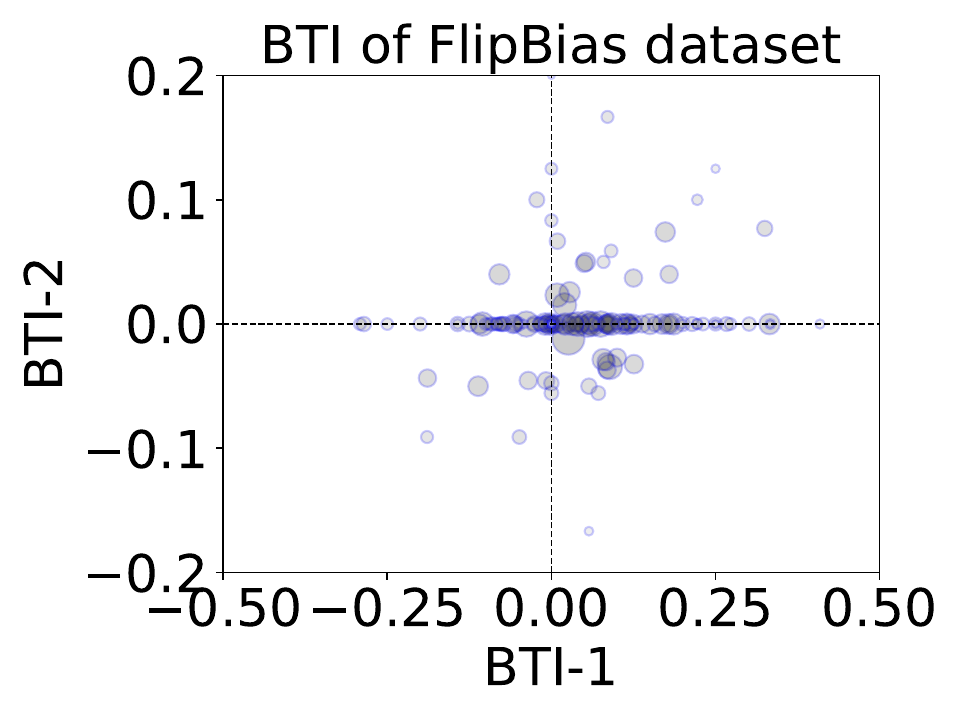}
}
\subfigure[12-Shot] {\label{sfig:appendix_FB_debias_topic_v1_v2_12shot}
\includegraphics[width=0.25\linewidth]{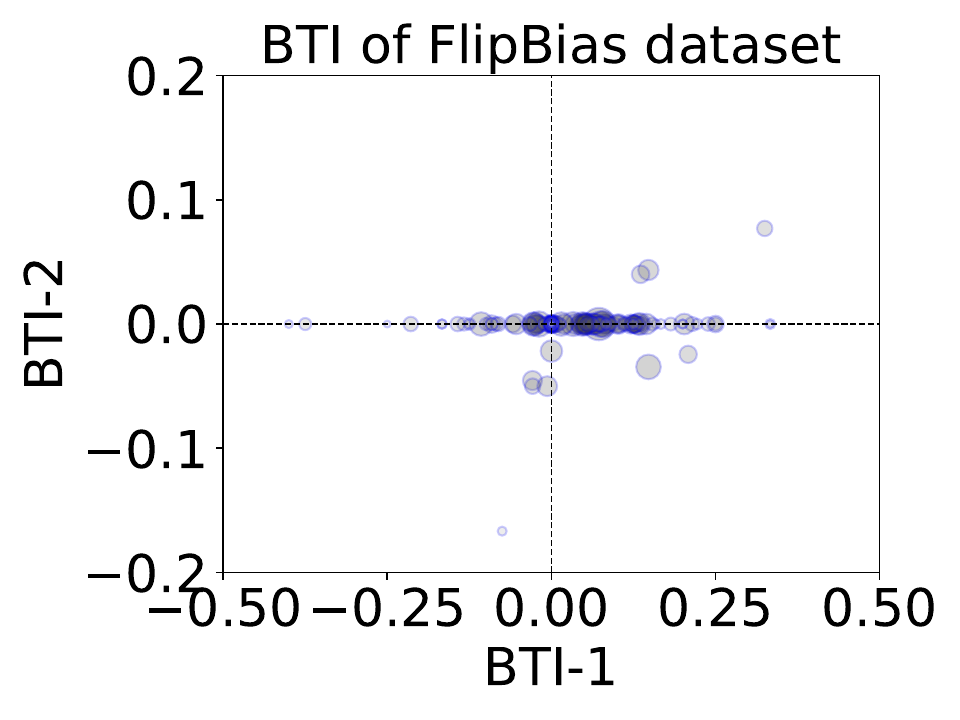}
}
\subfigure[DS] {\label{sfig:appendix_FB_debias_topic_v1_v2_DS}
\includegraphics[width=0.25\linewidth]{latex/pictures/debias_topic_FB_v1-v2_ds.pdf}
}
\subfigure[L-FT] {\label{sfig:appendix_FB_debias_topic_v1_v2_lft}
\includegraphics[width=0.25\linewidth]{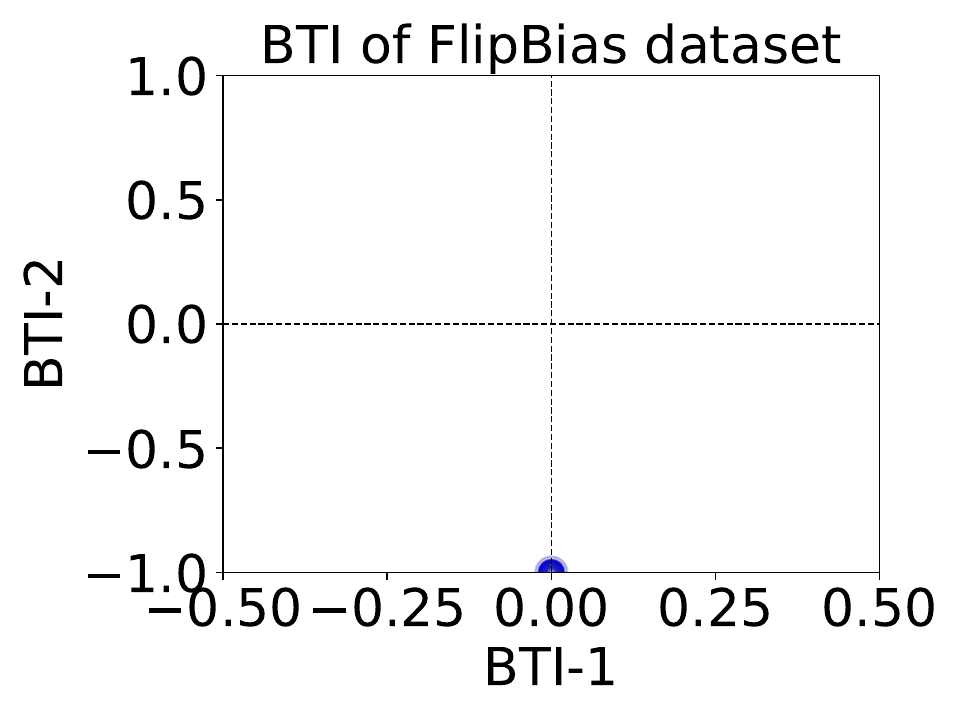}
}
\subfigure[LC-FT] {\label{sfig:appendix_FB_debias_topic_v1_v2_lcft}
\includegraphics[width=0.25\linewidth]{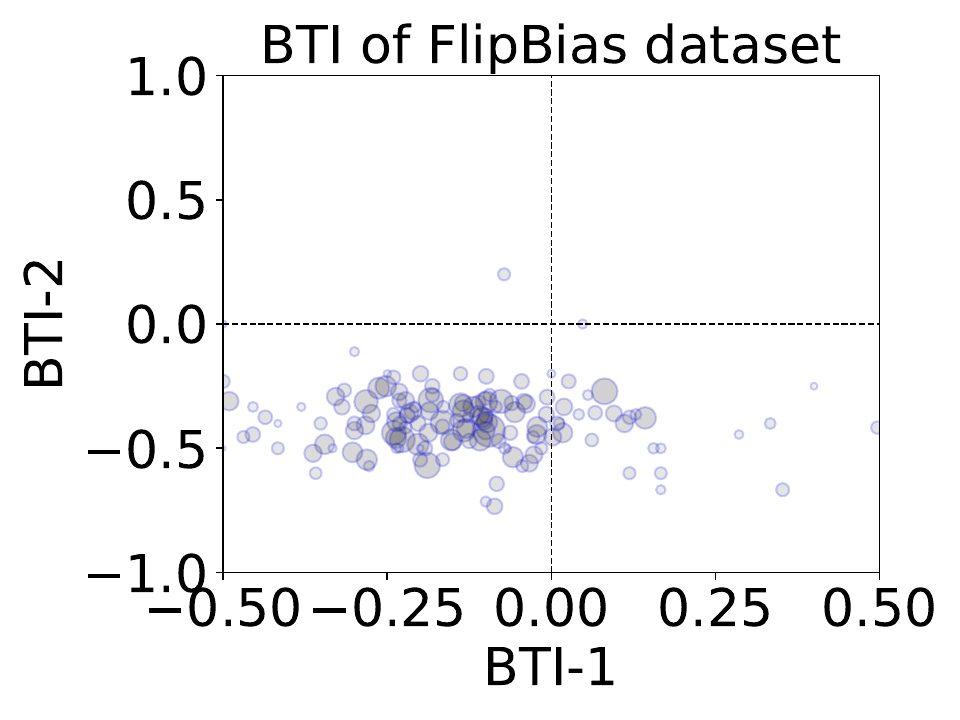}
}
\subfigure[LCR-FT] {\label{sfig:appendix_FB_debias_topic_v1_v2_lcrft}
\includegraphics[width=0.25\linewidth]{latex/pictures/debias_topic_FB_v1-v2_ftlcr.pdf}
}
\vskip -1em
\caption{\label{fig:appendix_FB_debias_topic_v1_v2} BTI distribution of Topics on FlipBias dataset after debiasing.
}
\vskip -1em
\end{figure*}

We list the BTI distribution of Topics on ABP and FlipBias datasets after prompt debiasing in \Cref{fig:ABP_debias_topic_v1_v2} and \Cref{fig:appendix_FB_debias_topic_v1_v2}, separately. 

\section{More Discussions}

\subsection{Distinguish from Related Works}

Existing research has explored political bias in LLMs. Here, we differentiate our contributions from key related works:

\cite{taubenfeld2024systematic} examines LLMs in simulating political debates, revealing conformity to inherent social biases despite specific directions. \cite{taubenfeld2024systematic}focuses on interaction simulation, whereas our research centers on bias detection, emphasizing end-to-end and fine-grained analyses. 
\cite{rozado2024political} analyzes bias through 11 political orientation tests, while our study highlights limitations of orientation tests and provides robust quantitative analysis based on extensive datasets, offering a broader perspective on LLM biases. 
\cite{urman2023silence} investigates LLM-Chat Models' responses to predefined queries, focusing on non-responses and false responses. While related to the political domain, it primarily addresses jailbreaking and harmful effects. Our research questions are more specific, targeting systematic bias detection. \cite{motoki2024more} evaluates ChatGPT's responses to ideological questions. It focuses solely on ChatGPT, whereas our work encompasses a broader range of LLMs and addresses comprehensive research questions, providing a more extensive analysis.

These studies contribute to understanding political bias in LLMs. However, our work stands out by offering a more systematic exploration, addressing four comprehensive research questions, employing intricate experimental designs, and analyzing a broader range of LLMs, thus significantly extending the current body of research.

\subsection{Exploration of Different Embeddings}

In Section \ref{ssec:article_continuation}, we explored the embedding-based similarity matching method using embeddings from the GPT-3.5 model. Here, we extend our investigation to include another embedding source: sentence-t5-base\footnote{\url{https://huggingface.co/sentence-transformers/sentence-t5-base}} (T5-Base). The continuation results using T5-Base embeddings are summarized in Table \ref{tab:appendix_continuation_embedding}. The calculation of left and right percentages in the table follows the methodology detailed in Figure \ref{fig:article_contiuation_results}.

From \Cref{tab:appendix_continuation_embedding}, we observe similar trends across different prefix lengths as shown in \Cref{fig:article_contiuation_results}, although there are slight variations in predictions for prefix length = 320. Overall, the findings indicate a predominant left-leaning trend in continued articles, consistent with our earlier observations using GPT-3.5 embeddings.

\subsection{Result of Article Continuation based on classifier}
\label{appendix:ssec:classifier}

\begin{table*}[ht]
\setlength{\tabcolsep}{1.1mm}\small
\newcommand{\tabincell}[2]{\begin{tabular}{@{}#1@{}}#2\end{tabular}}
\begin{center}
\begin{tabular}{l|ccccc}
\toprule
 & \tabincell{c}{20 \\ (left/right)} & \tabincell{c}{40 \\ (left/right)} & \tabincell{c}{80 \\ (left/right)} & \tabincell{c}{160 \\ (left/right)} & \tabincell{c}{320 \\ (left/right)} \\
\midrule
ZERO-SHOT & 41.8/58.2 & 19.3/80.7 & 20.0/80.0 & 41.1/58.9 & 27.3/72.7 \\
FEW-SHOT & 70.5/29.5 & 53.6/46.4 & 53.3/46.7 & 51.3/48.7 & 48.0/52.0 \\
\bottomrule
\end{tabular}
\end{center}
\caption{
Results of article continuation based on
classifier (GPT 3.5). 
}
\label{tab:appendix_claissifer} 
\end{table*}

In this experiment using \texttt{gpt-3.5-turbo-0613} as a classifier, we have two settings. In the setting zero-shot, we directly prompt ChatGPT3.5 to answer whether the given continuation is Left, Center, or Right, in the setting few-shot setting, we give the same reference as we compared in embedding-based similarity matching, which is a triple of Left-Center-Right articles that are of the same event as our continuation. The results of the ratio Left/Right are shown in \Cref{tab:appendix_claissifer}.

\subsection{Article Continuation Examples}
\begin{table*}[t]
\setlength{\tabcolsep}{1mm}\small
\begin{center}
\begin{tabular}{p{15cm}}
\toprule[1.0pt]
\textbf{Prefix-20 Example:} \\
Prefix: WASHINGTON (Reuters) - President Donald Trump and his Republican allies in Congress on Friday escalated a campaign against \\
\textcolor{brown}{Suffix}: Special Counsel Robert Mueller and the ongoing Russia probe, as an explosive memo was released that alleges misconduct within the FBI and Justice Department. The memo, written by Republican lawmakers, accuses the FBI of abusing its surveillance powers in obtaining a warrant to spy on a former Trump campaign aide… \\
\midrule
\textbf{Prefix-40 Example:} \\
Prefix: WASHINGTON (Reuters) - President Donald Trump and his Republican allies in Congress on Friday escalated a campaign against U.S. law enforcement agencies over their probe of Trump’s ties to Russia, releasing a disputed memo that \\
\textcolor{brown}{Suffix}: accuses the FBI of bias and misconduct. The explosive memo, which was released by the House Intelligence Committee, has sparked intense controversy and further deepened the divide between the Trump administration and the intelligence community. …  \\
\midrule
\textbf{Prefix-80 Example:} \\
Prefix: … Trump’s ties to Russia, releasing a disputed memo that the FBI warned was misleading and inaccurate. The previously classified document, written by Republicans on the House of Representatives intelligence committee, alleges that the federal probe of potential collusion between Trump’s 2016 presidential campaign and \\
\textcolor{brown}{Suffix}: Russian officials was politically motivated and based on questionable evidence. The release of the memo marks a significant turning point in the ongoing battle between Trump and the intelligence community. Trump has long criticized the investigation into Russian interference in the 2016 election, calling it a "witch hunt" and claiming that it is a politically motivated attempt to undermine his presidency. … \\
\bottomrule[1.0pt]
\end{tabular}
\end{center}
\caption{Examples of Article Continuation}
\label{tab:appendix_continuation_examples}
\end{table*}

\begin{table*}[ht]
\setlength{\tabcolsep}{1.1mm}\small
\newcommand{\tabincell}[2]{\begin{tabular}{@{}#1@{}}#2\end{tabular}}
\begin{center}
\begin{tabular}{l|ccccc}
\toprule
 & \tabincell{c}{20 \\ (left/right)} & \tabincell{c}{40 \\ (left/right)} & \tabincell{c}{80 \\ (left/right)} & \tabincell{c}{160 \\ (left/right)} & \tabincell{c}{320 \\ (left/right)} \\
\midrule
GPT3.5 & 52.8/47.2 & 53.1/46.8 & 54.4/45.6 & 54.8/45.2 & 41.3/58.7 \\
T5-base & 50.0/50.0 & 49.9/50.1 & 52.8/47.2 & 51.7/48.3 & 57.6/42.4 \\
\bottomrule
\end{tabular}
\end{center}
\caption{Comparison of Two Embeddings (GPT3.5 v.s. T5-Base) Results in Article Continuation Experiments}
\label{tab:appendix_continuation_embedding}
\end{table*}

In \Cref{ssec:article_continuation}, we adopt GPT-3.5 to conduct article continuation. We first report the average suffix length for each setting as follows: 490.1, 487.5, 479.0, 463.7, and 473.9 for prefixes with lengths of 20, 40, …, 320, respectively. Due to the strong capability of GPT-3.5, the generated suffixes are quite consistent with the prefix. \Cref{tab:appendix_continuation_examples} shows the randomly selected examples in different prefix settings of article continuations.

\subsection{Finetuning Other LLMs}

In addition to the finetuning debiasing results of \texttt{ChatGPT 3.5} reported in \Cref{tab:debias_perfromance}, we examined the finetuning debiasing method on a smaller LLM, specifically \texttt{LLaMa-2-7B-Chat}.

We report the BTI-1 and BTI-2 scores for LLaMa2 in \Cref{tab:llama2_scores}, where:
\begin{itemize}
    \item \textbf{LLaMa2:} \texttt{LLaMa-2-7B-Chat} without finetuning.
    \item \textbf{LLaMa2 LCR-FT:} \texttt{LLaMa-2-7B-Chat} finetuned according to the setting described in Section 5.2, with 300 articles evenly distributed among left-label, center-label, and right-label categories (LCR-FT).
    \item \textbf{LLaMa2 Finetune (Right Leaning data):} \texttt{LLaMa-2-7B-Chat} finetuned with 300 right-leaning data, where grounded center articles are labeled as left, and grounded right articles are labeled as center.
\end{itemize}
\begin{figure*}[ht]
\centering
\subfigure[LlaMa2] {\label{sfig:emnlp_llama2_topic_FB_v1-v2}
\includegraphics[width=0.25\linewidth]{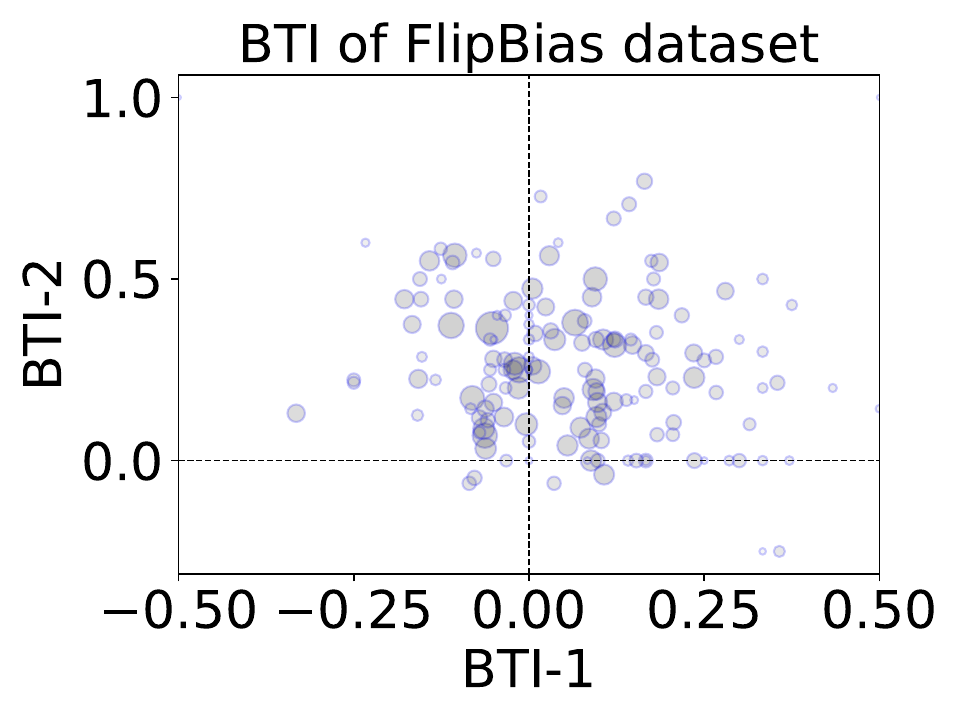}
}
\subfigure[LlaMa2 LCR-FT] {\label{sfig:emnlp_llama2_ft300_topic_FB_v1-v2}
\includegraphics[width=0.25\linewidth]{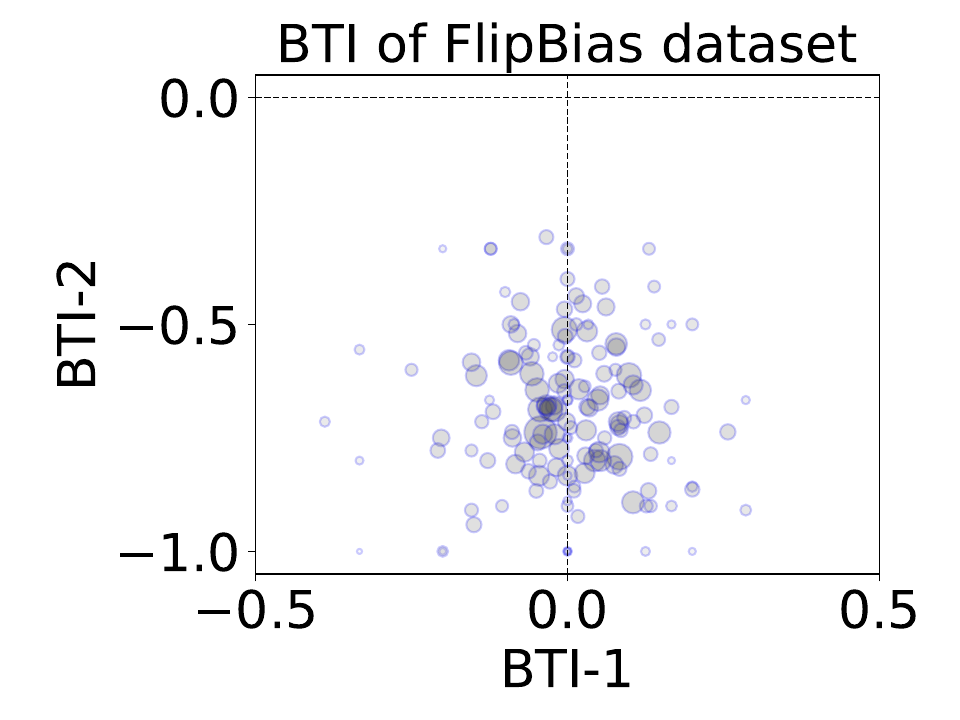}
}
\caption{\label{fig:appendix_llama2_finetune} BTI distribution of Topics on FlipBias dataset for \texttt{LlaMa2} and \texttt{LlaMa2 LCR-FT}.
}
\end{figure*}
We further report the BTI distribution of \texttt{LLaMa2} and \texttt{LLaMa2 LCR-FT} in \Cref{fig:appendix_llama2_finetune}. We observe that although the averaged BTI-1 scores do not exhibit significant changes in \Cref{tab:llama2_scores} before and after finetuning, upon examining the topic-level distribution (refer to \Cref{fig:appendix_llama2_finetune}), we notice a more centralized BTI-1 distribution.

\begin{table*}[ht]
\setlength{\tabcolsep}{1mm}\small
\begin{center}
\begin{tabular}{lcc}
\toprule[1.0pt]
Model & BTI-1 \\
\midrule
LLaMa2 & 0.04  \\
LLaMa2 LCR-FT & -0.024 \\
LLaMa2 Finetune (Right Leaning data) & 0.02\\
\bottomrule[1.0pt]
\end{tabular}
\end{center}
\caption{BTI-1 of LLaMa2 and Finetuning Methods}
\label{tab:llama2_scores}
\end{table*}

\end{document}